\documentclass[twocolumn]{aastex631}
\usepackage{appendix}
\usepackage{amsmath}
\usepackage{booktabs}
\usepackage{tablefootnote}
\usepackage{color}
\usepackage{xcolor}
\usepackage{subfigure}
\usepackage{multirow}
\graphicspath{{./}{figures/}}

\shorttitle{2DFIS Overview}
\shortauthors{Liu et al.}

\makeatletter
\def\endacknowledgements{%
  \expandafter\ifx\csname internallinenumbers\endcsname\relax\else
    \end{internallinenumbers}%
  \fi
  \egroup%% completes ackbox
  \ifanonymous
    \vskip 5.8mm plus 1mm minus 1mm
    \vskip1sp
    \centerline{(Acknowledgements anonymized for review)}%
  \else
    \unvbox\ackbox % allow page/column breaks (do NOT re-vbox it)
  \fi
  \vskip6pt%
}
\makeatother
\begin{document}
%%%%%%%%%%%%%%%%%%%%%%%%%%%%%%%%%%%%%%%%%%%%%%%%%%%%%%%%%%%%%%%%%%%%%%%%%%

\title{CFHT MegaCam Two Deep Fields Imaging Survey (2DFIS) I: Overview}

\correspondingauthor{Binyang Liu, Wentao Luo}
\email{byliu@pmo.ac.cn, wtluo@ustc.edu.cn}

\author[0000-0002-0561-7937]{Binyang Liu}
\affiliation{Purple Mountain Observatory, Chinese Academy of Sciences, Nanjing, Jiangsu 210008, China}

\author[0000-0003-1297-6142]{Wentao Luo}
\affiliation{School of Aerospace Information and Technology, Hefei Institute of Technology, Hefei, Anhui 230031, China}
\affiliation{Department of Astronomy, School of Physical Sciences,
University of Science and Technology of China, Hefei, Anhui 230026, China}

\author[0000-0001-9513-7138]{Martin Kilbinger}
\affiliation{Astrophysicist at CEA Paris-Saclay, Gif-sur-Yvette 84294, France}

\author[0000-0001-5422-1958]{Shenming Fu}
\affiliation{Kavli Institute for Particle Astrophysics and Cosmology, 2575 Sand Hill Road, Menlo Park, CA 94025, USA}

\author[0000-0003-0751-7312]{Ian Dell'Antonio}
\affiliation{Department of Physics, Brown University, Providence, RI 02912, USA}

\author[0000-0001-5356-2419]{Liping Fu}
\affiliation{Key Lab for Astrophysics, Shanghai Normal University, Shanghai 200234, China}

\author[0000-0003-3728-9912]{Xian Zhong Zheng}
\affiliation{Tsung-Dao Lee Institute and State Key Laboratory of Dark Matter Physics, Shanghai Jiao Tong University, Shanghai, 201210, China}

\author[0000-0002-4911-6990]{Yi-fu Cai}
\affiliation{Department of Astronomy, School of Physical Sciences,
 University of Science and Technology of China, Hefei, Anhui 230026, China}

\author{Cheng Jia}
\affiliation{Key Lab for Astrophysics, Shanghai Normal University, Shanghai 200234, China}

\author[0000-0002-7152-3621]{Ning Jiang}
\affiliation{Department of Astronomy, School of Physical Sciences,
University of Science and Technology of China, Hefei, Anhui 230026, China}

\author[0000-0003-3616-6486]{Qinxun Li}
\affiliation{Department of Physics and Astronomy, The University of Utah, 115 South 1400 East, Salt Lake City, UT 84112, USA}

\author{Yicheng Li}
\affiliation{Key Lab for Astrophysics, Shanghai Normal University, Shanghai 200234, China}

\author[0009-0000-5381-7039]{Shurui Lin}
\affiliation{Department of Astronomy, University of Illinois at Urbana-Champaign, 1002 West Green Street, Urbana, IL 61801, USA}

\author[0000-0002-1365-0841]{Christopher J. Miller}
\affiliation{Department of Astronomy, University of Michigan, Ann Arbor, MI 48109, USA}
\affiliation{Department of Physics, University of Michigan, Ann Arbor, MI 48109, USA}

\author[0000-0002-2986-2371]{Surhud S. More}
\affiliation{The Inter-University Centre for Astronomy and Astrophysics, Ganeshkhind, Pune 411007, India}

\author[0000-0002-4911-6990]{Huiyuan Wang}
\affiliation{Department of Astronomy, School of Physical Sciences,
University of Science and Technology of China, Hefei, Anhui 230026, China}

\author{Yibo Wang}
\affiliation{Department of Astronomy, School of Physical Sciences,
University of Science and Technology of China, Hefei, Anhui 230026, China}

%%%%%%%%%%%%%%%%%%%%%%%%%%%%%%%%%%%%%%%%%%%%%%%%%%%%%%%%%%%%%%%%%%%%%%%%%%
\begin{abstract}
% We present the Two Deep Fields Imaging Survey (2DFIS), a program utilizing the Canada-France-Hawaii Telescope (CFHT) to study two distinct regions of the sky: one containing a repeating fast radio burst (FRB) source and the other featuring a dynamically ``rotating'' galaxy cluster.
% Reaching an r-band depth of 26th magnitude, the survey investigates the dark environment surrounding the FRB to search for potential optical counterparts.
% Simultaneously, detailed photometric and shape analyses of the galaxy cluster field enable a weak gravitational lensing study of its mass distribution. 
% This paper describes the observational strategies and data processing techniques used, including the LSST Science Pipelines with tailored modifications for CFHT data and the specific requirements of the survey.
% We provide a comprehensive overview of the workflow for processing CFHT imaging data, yielding key outputs such as calibrated single-epoch exposures, multi-band coadded images, and extensive source catalogs.
% These products form the foundation for future weak lensing and cosmological study.
%
We present the Two Deep Fields Imaging Survey (2DFIS), a wide-field imaging program conducted with the Canada–France–Hawaii Telescope (CFHT) targeting two astrophysically distinct regions: one containing a repeating fast radio burst (FRB) source and another hosting a candidate of a rotating galaxy cluster. 
Achieving a depth of $r \sim 26$~mag, the survey enables a search for faint optical counterparts and environmental signatures associated with the FRB, while high-quality photometric and galaxy shape measurements in the cluster field support a weak-lensing analysis of its mass distribution.
This paper describes the observing strategy and data processing methodology adopted for 2DFIS, including the use of the LSST Science Pipelines with survey-specific adaptations for CFHT/MegaCam data. 
We outline a complete workflow for transforming raw CFHT exposures into science-ready data products, including calibrated single-epoch images, multi-band coadded mosaics, and extensive source catalogs. 
These data products provide the foundation for ongoing and future studies of FRB host environments, cluster mass reconstruction, and related cosmological applications.

\end{abstract}
%%%%%%%%%%%%%%%%%%%%%%%%%%%%%%%%%%%%%%%%%%%%%%%%%%%%%%%%%%%%%%%%%%%%%%%%%%

\keywords{Wide field surveys; Astronomy data analysis; Weak gravitational lensing; Dark matter}

\section{Introduction}
\label{sec:intro}

Deep multi-band imaging surveys not only greatly deepen our understanding of clusters of galaxies, but also act as a gold mine for new discoveries. The deep imaging data from HST/ACS targeting the bullet cluster \citep{Clowe2006ApJ} provided the ``smoking gun'' evidence of the existence of dark matter, where the X-ray emission significantly deviates from the spatial distribution of dark matter inferred from weak gravitational lensing. 
The controversial ``dark core'' of the Abell 520 cluster, also known as the ``train wreck cluster'' \citep{mahdavi2007ApJ}, was discovered by deep imaging observations based on the 3.6-meter Canada-France-Hawaii-Telescope (CFHT) with the MegaCam instrument \citep{Boulade2003SPIE}, while \cite{Clowe2012ApJ} claim that the dark core disappears in their analysis. 
% The Local Volume Complete Cluster Survey (LoVoCCS) cluster survey \citep{Fu2022ApJ,Fu2024arXiv240210337F} applies the Dark Energy Camera (DECam) installed on the 4-meter Blanco telescope on the summit of Mt. Cerro Tololo. It provides weak lensing mass estimation for hundreds of clusters with extra X-ray detection. 
More recently, the Local Volume Complete Cluster Survey (LoVoCCS; \citealt{Fu2022ApJ,Fu_2024}) has utilized DECam on the 4-meter Blanco Telescope to obtain weak-lensing mass measurements for hundreds of nearby X-ray luminous clusters. 
It also serves as the low-redshift anchor for the Rubin Observatory Legacy Survey of Space and Time (LSST), which has recently begun its 10-year survey of the southern sky to explore dark energy, dark matter, and a wide range of astrophysical phenomena \citep{Ivezic_2019}. 
Together, these surveys highlight the unique power of deep optical imaging for probing cluster assembly, mapping dark matter, and identifying unusual dynamical features. 

% The discovery of dark matter deficit ultra defuse galaxies \citep{dokkum2018Natur,dokkum2019ApJ}, facilitated by the deep imaging from eight 400mm Dragonfly telephoto array telescopes \citep{Abraham2014PASP}. 
Deep imaging observations have also enabled the discovery of dark-matter–deficient ultra-diffuse galaxies (UDGs) \citep{dokkum2018Natur,dokkum2019ApJ}, made possible by the low-surface-brightness (LSB) sensitivity of the Dragonfly array of eight 400mm telephoto lenses \citep{Abraham2014PASP}. 
These observations sparked a number of theories of the formation of these objects, e.g. high-velocity galaxy collisions \citep{Silk2019MNRAS} or the mini-bullet cluster scenario \citep{dokkum2022Natur}. 
The search for extremely large LSB galaxies or UDGs \citep{Bautista2023ApJS} in the Coma Cluster using Subaru HSC and the Next Generation Virgo Cluster Survey (NGVS) with CFHT \citep{Ferrarese2012ApJS} provides valuable information on the formation mechanism of UDGs. 

Among ground-based optical facilities, the CFHT MegaCam stands out with its excellent imaging quality at the observation site at Maunakea, and has become a valuable resource for many scientific fields. 
The CFHTLS and, subsequently, the CFHTLenS project \citep{Hoekstra_2006,Fu_2008,Heymans2012MNRAS} were pioneering surveys focusing on weak gravitational lensing using observations that started in 2003. They can be placed after galaxy-galaxy lensing studies from SDSS DR4 \citep{sheldon2004AJ,mandelbaum2006MNRAS} and before SDSS DR7 \citep{luo2017I,luo2018II}. 
The deep imaging resulted in a very high galaxy number density ($\sim18$ galaxies~arcmin$^{-2}$) enabled high signal-to-noise ratio (SNR) measurements of lensing convergence maps \citep{VanWaerbeke2013MNRAS} as well as the second-order shear-shear correlation function \citep[cosmic shear;][]{Fu2014MNRAS}, which is a probe of the matter-matter power spectrum that can be used to constrain cosmology \citep{Kilbinger2013MNRAS}. 
A more recent project, the Ultraviolet Near Infrared Optical Northern Survey \citep[UNIONS;][]{Gwyn_2025}, aiming to provide multi-band photometry for the Euclid space telescope, is using CFHT in the $u$ and $r$ bands in addition to the $i$ band from Pan-STARRS and $g$ and $z$ bands from Subaru. 
A recent work using UNIONS weak-lensing data probed the scaling relation between supermassive black holes (SMHB) and the dark matter halo \citep{Li2024arXiv240210740L}. This is the first direct measurement of this 
a tight relationship between small-scale SMBHs and their large-scale dark matter halos.

Fast radio bursts (FRBs) remain a mystery since their discovery in 2007 \citep{Lorimer2007Sci}. 
Although their formation mechanism is uncertain, most of them are known to be extra-galactic sources. 
So far, more than 500 FRBs have been detected with the Canadian Hydrogen Intensity Mapping Experiment (CHIME; \citealt{CHIME2022ApJS}) alone. 
An increasing number of FRBs, in particular those with accurate position measurements, can be matched with their host galaxies from optical observations \citep{Bhandari2022AJ,Law2024ApJ}, which initiated further study of FRB host galaxies properties.

In this work, we have observed two sky regions with the CFHT MegaCam wide-field imager. 
One is targeting a dynamically active cluster (RXCJ0110.0+1358) selected from the SDSS DR7 group catalog \citep{Yang2007ApJ}. 
The reason we select this specific cluster is that it shows indications of rotation or merger features, with two systematically blue/red-shifted parts divided by an axis. In addition, the cluster's satellite distribution is not symmetric.

The second field is targeting the FRB repeater (three bursts in 2019) FRB190417 from the CHIME catalog. The time difference among the FRB repeating events are from tens of days to hundreds days, which is similar to a multiple-imaged strong lensing system. We try to find an optical strong lensing system in this field and to study if this repeater is caused by strong lensing like a ``Refsdal system" or not.

We carry out deep imaging in $ugri$ to map the dark environment of these two objects. We name the combination of the two separate areas the Two Deep Fields Imaging Survey (2DFIS). This paper is organized as follows. 
We detail the survey design in Section~\ref{sec:observation} including the target fields and specific observing strategy. 
In Section~\ref{sec:data_processing}, we address the data processing procedure with the LSST Science Pipelines and the LoVoCCS pipeline to process the raw data and generate the catalogs. 
We show the overall results and quality control in Section~\ref{sec:data} and further discuss the potential systematics in Section~\ref{sec:discussion}. We conclude in Section~\ref{sec:conclusions}. 
\section{Observation}
\label{sec:observation}

\subsection{Survey Field}
\label{sec:survey_field}

The Two Deep Fields Imaging Survey (2DFIS) consists of two distinct CFHT MegaCam fields designed to probe complementary astrophysical environments. 
The first field, designated \textit{Cluster Rot~1} or RXCJ0110.0+1358, is centered at $\alpha = 01^{\mathrm h}10^{\mathrm m}00^{\mathrm s}$, $\delta = +13^\circ58'00''$ (J2000). 
It corresponds to a nearby ($z \simeq 0.058$) galaxy cluster identified as dynamically ``rotating'' or merging along the line of sight.  
The field was selected from the SDSS~DR7 group catalog \citep{Yang2007ApJ} for its high halo mass, $\log (M_{200}/M_\odot) = 14.42$, and distinctive azimuthal velocity pattern of $\sim600\,\mathrm{km\,s^{-1}}$, making it a promising laboratory for testing the two-dimensional weak-lensing reconstruction of mass substructures and for exploring the interplay between baryons and dark matter in non-relaxed systems. 

The second field targets the repeating fast radio burst (FRB) source FRB190417 \citep{CHIME_2019}, located at $\alpha = 17^{\mathrm h}05^{\mathrm m}00^{\mathrm s}$, $\delta = +68^\circ17'00''$ (J2000).  
This high-declination field is designed to characterize the dark and baryonic environment surrounding the FRB, enabling a search for potential optical counterparts and host candidates through deep multi-band imaging.  
The FRB region serves as a contrasting case to the cluster field, tracing isolated extragalactic structures and diffuse matter under different environments.

% Together, these two fields form the foundation of 2DFIS, providing an opportunity to bridge small-scale baryonic phenomena with large-scale dark matter mapping.

\subsection{Observing Strategy}
\label{sec:obs_strategy}

Both observations were conducted with the MegaCam imager installed at the CFHT focal plane.  
The cluster field (Run~ID 22BD07) and the FRB field (Run~ID 22BS10) were observed under photometric or near-photometric conditions during the 2022B semester.  
Both fields were imaged in the $u$, $g$, $r$ and $i$ bands, following a consistent dither pattern to fill chip gaps and improve sky flatness.  
The standard CFHT pattern DP4 was adopted with four positions per filter, scaled by a factor of 1.5 relative to the nominal offsets.

For the rotating cluster RXCJ0110.0+1358 field, the total integration time was 3.8~hr, distributed as $4\times960$~s ($u$), $4\times420$~s ($g$), $4\times920$~s ($r$), and $4\times960$~s ($i$), with the limiting magnitude of $\sim26.0$ by design for all bands. 
Observations were conducted in gray time with an airmass limit of $<1.2$, and image quality constrained between 0.65\arcsec\ and 0.80\arcsec.  
The uniform seeing and multi-band coverage were designed to enable accurate color measurements, while the weak-lensing analysis and mass map reconstruction are performed using the deep $r$-band imaging, which provides the highest signal-to-noise ratio and most stable point spread function (PSF) for galaxy shape measurements. 

For the FRB190417 field, the total exposure time was 3.4~hr, distributed as $4\times900$~s ($u$), $4\times200$~s ($g$), $4\times600$~s ($r$), and $4\times650$~s ($i$).  
Observations were executed under dark or gray sky conditions, with image quality requirements of 0.65–0.80\arcsec\ in the $r$ band and airmass below 2.0.  
The relatively high declination of the target resulted in elevated airmass values, which were accounted for in the final exposure plan to reach a depth of $i_{\mathrm{AB}} \simeq 25.2$~mag.

A summary of the observational parameters, including coordinates, filter configurations, exposure times, and observing constraints, is provided in Table~\ref{tab:obs}.

\subsection{Instrumentation}
\label{sec:instrumentation}

The survey employed the MegaCam wide-field optical imager mounted at the prime focus of CFHT.  
MegaCam consists of 36 CCDs, each $2048\times4612$~pixels, providing a contiguous field of view of $1\deg\times1\deg$ with a pixel scale of $0.187''\,\mathrm{pixel^{-1}}$.  
The instrument delivers excellent image quality over the full focal plane, making it particularly well suited for deep extragalactic imaging and weak-lensing applications.  

All filters used in 2DFIS belong to the standard CFHT MegaCam photometric system: $u.MP9302$, $g.MP9402$, $r.MP9602$, and $i.MP9703$.  
Guiding was performed in sidereal mode for all exposures.  
The combination of high-quality optics, stable atmospheric conditions at Maunakea, and consistent dithered coverage ensures homogeneous photometric calibration across the 1~deg$^2$ field.  
The resulting data reach an $r$-band depth of $\sim26$~mag (5$\sigma$, point source), enabling robust detection of faint background galaxies and diffuse structures around both targets.

\begin{deluxetable*}{lccccccc}
\tablecaption{Summary of CFHT/MegaCam Observations in 2DFIS\label{tab:obs}}
\tablehead{
\colhead{Field} & \colhead{Run ID} & \colhead{R.A. (J2000)} & \colhead{Decl. (J2000)} & 
\colhead{Filters} & \colhead{Exp. Time (s)} & \colhead{Seeing Requirement (\arcsec)} & \colhead{Airmass Limit}
}
\startdata
RXCJ0110.0+1358 & 22BD07 & 01:10:00 & +13:58:00 & $u,g,r,i$ & 960, 420, 920, 960 & 0.65–0.80 & $<1.2$ \\
FRB~20181017A & 22BS10 & 17:05:00 & +68:17:00 & $u,g,r,i$ & 900, 200, 600, 650 & 0.65–0.80 & $<2.0$ \\
\enddata
\tablecomments{All observations used the DP4 dither pattern with four exposures per filter and a pattern scale factor of 1.5. Image quality refers to the $r$-band seeing requirement at half-light radius.}
\end{deluxetable*}
\section{Data Processing}
\label{sec:data_processing}

The LSST Science Pipelines provide a flexible and comprehensive data-management framework designed to accommodate diverse imaging datasets and support a broad range of scientific goals. They incorporate state-of-the-art algorithms for image processing and source measurement, reflecting the latest methodological advances in astronomical data analysis.

In this work, the observational data from both fields are processed using the LSST Science Pipelines. 
Although originally developed for LSST, the pipelines are explicitly designed to support data from other facilities. 
Their algorithmic development builds on expertise accumulated from previous large surveys—including SDSS \citep{York_2000,Lupton_2001}, Pan-STARRS \citep{Magnier_2020,Kaiser_2010}, DES \citep{Sevilla_2011,Morganson_2018}, SuperMACHO \citep{Becker_2005}, DLS \citep{Wittman_2002}, and CFHTLS \citep{Erben_2013,Kilbinger_2013,Gwyn_2008} — as well as on established software tools such as SExtractor \citep{Bertin_SExtractor_1996} and PSFEx \citep{Bertin_PSFEx_2013}. 
The LSST Science Pipelines also serve as the primary reduction system for Hyper Suprime-Cam (HSC) on the Subaru Telescope and have enabled two public data releases \citep{Miyatake_2012,Bosch_2017,Li_2022}.
As an open-source software stack, the pipelines expose a Python interface for high-level user interaction, while performance-critical components are implemented in C++ and interfaced through SWIG wrappers \citep{beazley1996swig}. 
The demonstrated success of the pipeline has shown that large astronomical datasets can be autonomously reduced, with reliable source detection and the extraction of scientifically meaningful measurements.

% In the next stage of dark energy surveys, we intend to expand the survey coverage to encompass additional sky regions while simultaneously exploring synergies with existing archival and ongoing survey data. 
% As a consequence, it becomes imperative to adopt a processing framework that ensures uniformity, utilizes consistent methods, and adheres to a compatible catalog structure. 
% Within this context, the LSST Science Pipelines emerges as the preferred solution, guaranteeing both data coherence and operational convenience. 
% An additional advantage lies in its well-established parallel processing mechanism, which significantly enhances efficiency by optimizing CPU utilization and minimizing time expenditure \textcolor{red}{(REF)}. 

In each targeted sky area, the raw CFHT images undergo detrending prior to executing astrometric and photometric calibration at the level of individual CCDs. 
Subsequently, these single-frame images are stacked to generate coadded images, wherein static-sky objects are identified and their properties quantified at the coadd level, as elaborated upon in Section~\ref{subsec:coaddition}. 
While the image-level processing through coadd measurements is performed using the LSST Science Pipelines, all analysis steps following coaddition measurements, including higher-level catalog refinement, and the generation of science-driven data products, are carried out within the LoVoCCS pipeline \citep{Fu2022ApJ,Fu_2024}. 
This separation is necessary to accommodate customized processing and analysis modules required for deep-field studies, such as optimized source selection, photometric redshift estimation, and weak-lensing–oriented measurements, which are beyond the scope of the standard LSST workflow.
The adoption of the LoVoCCS pipeline is well motivated, as the scientific objectives of the 2DFIS survey closely align with those of the LoVoCCS survey, particularly in weak lensing studies of galaxy clusters and large-scale dark matter distribution, and the two surveys share comparable image quality requirements and source selection criteria. 

A comprehensive suite of Quality Control (QC) tests is applied to all data products to validate both data quality and processing performance, as described in Section~\ref{subsec:qc}. 
The resulting catalogs, generated by the LSST Science Pipelines and subsequently refined through the LoVoCCS pipeline, provide the foundation for subsequent analyses, including photometric redshift estimation and cluster mass modeling.

In the following, we provide a concise overview of the processing stages utilizing the LSST Science Pipelines (version 19.0.0) and the LoVoCCS pipeline, incorporating configuration parameters specifically adapted for CFHT data. 
For a broader and more comprehensive insight into the commands and application of the LSST Science Pipelines, we recommend referring to the works of \cite{Bosch_2017,Bosch_2019,Ivezic_2019,Jenness_2022}. 

\begin{figure*}
    \centering
    \includegraphics[width=1\textwidth]{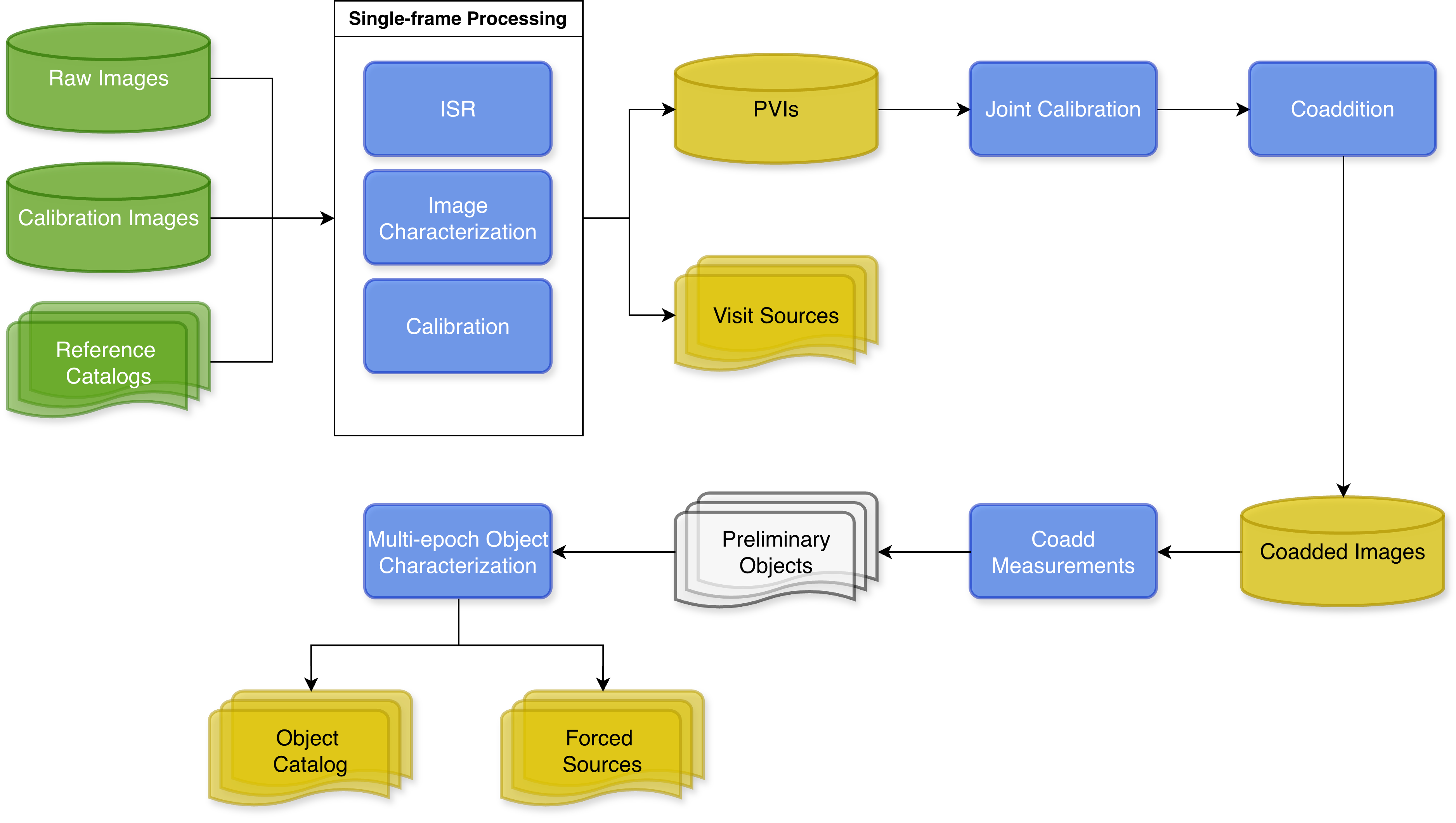}
    \caption{The 2DFIS data processing workflow}
    \label{fig:flow_chart_all}
\end{figure*}

In this section, we describe the data processing workflow including detrending, calibration, selection, coaddition, and measurement of the CFHT data. 
A flowchart is illustrated in Fig.~\ref{fig:flow_chart_all}.
Each sky field yields processing outputs of approximately 1.0 TB in size, requiring a computational time of approximately 1–2 days when employing 20 cores (Intel Skylake/Cascade architecture). 
The memory requirement stands at around 100 GB. 

%%%%%%%%%%%%%%%%%%%%%%%%%%%%%%%%%%%%%%%%%%%%%%%%%%%%%%%%%%%%%%
%%%%%%%%%%%%%%%%%%%%%%%%%%%%%%%%%%%%%%%%%%%%%%%%%%%%%%%%%%%%%%
\subsection{Data Initialization}
\label{subsec:data_init}

We begin by ingesting the raw images into the LSST \texttt{Butler} framework, along with the corresponding raw bias and flat frames. 
Additionally, we create repositories of reference catalogs, including data from SDSS, Gaia, and Pan-STARRS, for astrometric and photometric calibrations.

\subsubsection{Raw Image Ingestion}

We begin by creating a data repository to store all input images along with the intermediate data products generated throughout the subsequent processing steps. 
The raw images, together with all required metadata, are ingested into this repository. 
The underlying registry is implemented using SQLite and accessed via SQLAlchemy\footnote{\url{https://www.sqlalchemy.org}}. 
Only low-level SQLAlchemy interfaces are used for table creation and interaction \citep{Jenness_2022}. 
The SQLite registry is populated with metadata extracted from the raw image headers, including observation date, telescope pointing, detector and CCD identifiers, filter information, and other exposure-level attributes.

The LSST Science Pipelines Data Management System (the \texttt{Data Butler}) is designed to abstract away the complexities of data access. 
Users are not required to know telescope-specific parameters or instrument-dependent file formats.
Instead, they retrieve data using standardized astronomical concepts.

To support CFHT imaging, a \texttt{mapper} file is added to the repository to define the correspondence between CFHT CCDs and the LSST data framework. 
This mapper connects to the \texttt{obs\_cfht} package, which is integrated within the LSST Science Pipelines and enables the pipelines to ingest and process CFHT MegaCam data according to the appropriate instrument format, camera geometry, defect masks, and other instrument signatures.

\subsubsection{Reference Catalogs} 

Reference star catalogs required for both astrometric and photometric calibration are prepared and deployed at the beginning of the data-processing workflow. 
These catalogs are utilized during the \texttt{processCcd} and \texttt{jointcal} stages. 
For astrometric calibration, we adopt the \textit{Gaia} Data Release~2 (Gaia~DR2), which provides all-sky coverage and sub-milliarcsecond positional accuracy \citep{Gaia1_2018}. 

For photometric calibration, we account for the mismatch between the MegaCam filter system and that of the reference surveys. 
Due to the degeneracy between the \textit{Gaia} blue/red photometers and MegaCam bandpasses, we employ the Pan-STARRS1 Data Release~1 (PS1~DR1) as the photometric reference for the $g,r,i$ bands, and SDSS~DR12 for the $u$ band. 
The original reference catalogs, provided in the Hierarchical Triangular Mesh (HTM) format, are converted into the LSST-compatible reference catalog structure. 

Because of residual differences between the MegaCam and reference-system bandpasses, color-term corrections are applied following single-epoch processing to ensure consistent photometric calibration across filters.

%%%%%%%%%%%%%%%%%%%%%%%%%%%%%%%%%%%%%%%%%%%%%%%%%%%%%%%%%%%%%%
%%%%%%%%%%%%%%%%%%%%%%%%%%%%%%%%%%%%%%%%%%%%%%%%%%%%%%%%%%%%%%
\subsection{Single-Frame Processing}
\label{subsec:single_frame}

After ingesting the raw data into a \texttt{Butler} repository, single-frame data reduction is performed on each CCD. 
The corresponding user-level command is \texttt{processCcd.py} or \texttt{singleFrameDriver.py} which performs a series of tasks including instrumental signature removal, CCD image characterization, and CCD-level calibration.
% The procedures of single-frame processing are shown in Fig. ~\ref{fig:flow_chart_singleFrame}.

% \begin{figure*}
%     \centering
%     \includegraphics[width=1\textwidth]{figures/flow_chart_singleFrame.png}
%     \caption{Single-frame processing workflow}
%     \label{fig:flow_chart_singleFrame}
% \end{figure*}

\subsubsection{Instrumental Signature Removal (\texttt{ISR})}

Instrumental Signature Removal (ISR) comprises a sequence of corrections, including the assembly of multiple amplifier readouts into full CCD images, removal of overscan and non-linearity effects, mitigation of electronic crosstalk, and the application of flat-field, bias, and dark corrections, along with the construction of variance and mask planes. 
In our workflow, the raw CFHT images have already been pre-processed by the CFHT Elixir system\footnote{\url{https://www.cfht.hawaii.edu/Instruments/Elixir/overview.html}}, which performs image-quality assessment, calibration analysis, and metadata compilation. 
Because Elixir provides the bulk of the instrument detrending using master calibration data, no additional master calibration inputs are required during our ISR stage.

Saturated and defective pixels are identified, flagged, and subsequently interpolated. 
Pixel interpolation is performed using Gaussian Processes following the Linear Predictive Code (LPC) method \citep{Press_2007}, which estimates missing or corrupted pixel values from neighboring samples and their statistical properties. 
Mask planes are updated to record both the presence of interpolation and its underlying cause. 
Sources whose centroids fall on interpolated pixels are flagged as having unreliable measurements in the resulting catalogs. 
ISR-corrected exposures, together with their mask planes, are then written to the \texttt{Butler} repository. 

CFHT MegaCam utilizes thinned E2V Type~42-90 CCDs, for which the brighter–fatter effect has an amplitude of approximately $0.5\%$ of the full dynamic range \citep{Antilogus_2014}. 
This contribution is negligible compared to the intrinsic shape noise and has minimal impact on the photometry and shape measurements of individual objects. 
Accordingly, the brighter–fatter correction is disabled in our pipeline processing.

\subsubsection{Image Characterization (\texttt{CharImage})}

Starting from the ISR-corrected exposures (\texttt{postISRCCD}), bright sources are first detected and preliminarily measured. 
Cosmic rays are then identified and interpolated. 
A sky background model is constructed accounting for all mask-plane information, and subtracted from each CCD image. 
At this stage, the background subtraction is performed independently for each CCD. 
A second, focal-plane–level sky model is later derived during coaddition to mitigate background discontinuities across CCD boundaries.

Following background removal, the LSST Science Pipelines construct a PSF model for each CCD. 
Bright detected sources are initially classified as point-like or extended based on their preliminary photometric properties. 
PSF estimation is iteratively refined through two rounds of star selection, retaining only high-quality point sources. 
In the single-epoch catalogs, all candidate PSF stars are flagged as \texttt{PSF\_candidate=1}, while those ultimately used in the final PSF model carry the flag \texttt{PSF\_used=1}.

Each detected source is represented as a \texttt{footprint}. 
Initial aperture photometry is performed on these footprints, followed by aperture-correction measurements. 
After coaddition, in the subsequent \texttt{deblendCoaddSources} and \texttt{measureCoaddSources} stages, the coadd-level footprints are deblended and photometric measurements are recomputed for each deblended child object.

% detect and measure bright sources - repair cosmic rays - measure and subtract background - measure PSF 

%%%
Photometric measurements at both the single-epoch CCD level and the coadd level are carried out using the \texttt{CModel} algorithm, a refined successor of the model-fitting approach introduced in the \textit{SDSS Photo} pipeline \citep{Lupton_2001,ivezic2004sdss,lupton2005sdss}. 
\texttt{CModel} provides a robust and flexible means of estimating total fluxes by fitting PSF-convolved galaxy models in a controlled sequence, effectively approximating bulge-disk decomposition or S\'ersic-profile representation \citep{Sersic_1963}. 

The fitting procedure proceeds in three stages: an initial fit with an elliptical exponential model ($n=1$), followed by a de~Vaucouleurs model ($n=4$; \citealt{deVaucouleurs1948}), and finally a combined fit in which the structural parameters are fixed and only the amplitudes of the two components are allowed to vary \citep{Bosch_2017}. 
This staged strategy stabilizes the optimization, reduces parameter degeneracies, and yields reliable total-flux estimates even when the SNR is modest. 
Because all models are explicitly convolved with the locally estimated PSF, \texttt{CModel} captures the extended light profiles of galaxies while preserving accurate flux recovery for compact sources.

A key advantage of \texttt{CModel} photometry is its robustness on single-epoch images, where noise, seeing variations, and limited depth can constrain more complex structural modeling. 
The algorithm requires only a small number of well-constrained parameters at each stage, making it resistant to overfitting and well-suited for varied observing conditions. 
In particular, the PSF-convolved modeling ensures that the inferred total flux remains stable across a wide range of PSF sizes, which is critical when measuring sources on individual CCD exposures before coaddition.

For each detected object, both PSF and \texttt{CModel} magnitudes are measured. 
Their difference defines the \texttt{extendedness} parameter, which serves as a reliable morphological classifier. 
Stellar sources yield \texttt{extendedness} values near zero because the galaxy model collapses to a PSF-like profile, whereas resolved galaxies produce distinctly positive values due to the systematic underestimation of their fluxes by PSF photometry. 
This metric provides an efficient and physically motivated means of distinguishing point-like and extended sources in single-frame catalogs. 

While \texttt{CModel} is optimized for photometry rather than weak-lensing shear measurements, it delivers stable, PSF-corrected fluxes that are sufficiently precise for the downstream tasks in this work, including photometric-redshift estimation and cluster science based on coadded measurements. 
For the weak-lensing analyses presented later, galaxy shape measurements will instead be performed using dedicated shear-estimation algorithms such as \texttt{shapeHSM} \citep{Hirata_2003,Mandelbaum_2005}, \texttt{NGMIX} \citep{Sheldon_2015}, \texttt{lensfit} \citep{Miller_2013}, or \texttt{FourierQuad} \citep{Zhang_2022}.

\subsubsection{Calibration}

Astrometric and photometric calibration is performed after image characterization, once a stable PSF model and aperture correction map have been established for each exposure. 
As noted in Section~\ref{subsec:data_init}, we adopt Gaia~DR2 as the astrometric reference catalog \citep{Gaia_2016,Gaia_2023}, and use PS1~DR1 and SDSS~DR12 as the photometric reference catalogs \citep{Flewelling_2020,Chambers_2019,Alam_2015}. 
A preliminary deblending step is applied to the detected footprints, and well-measured, isolated sources are selected for cross-matching with the reference catalogs. 

Astrometric calibration is carried out iteratively, typically converging within three iterations. 
The resulting World Coordinate System (WCS) solution achieves a scatter of order $\sim 0.01\arcsec$, modeled using third-order Simple Imaging Polynomial (SIP) distortion terms. 

After refining the WCS, sources are positionally matched to the photometric reference catalogs. 
Although Gaia provides all-sky coverage and exceptionally accurate astrometry, its photometric system is limited to broad BP and RP passbands\footnote{\url{https://www.cosmos.esa.int/web/gaia/dr2}} and is not well suited for calibrating CFHT MegaCam filters. 
Using Gaia photometry alone would introduce strong degeneracies in color-term transformations between the Gaia and MegaCam systems. 
Moreover, no existing all-sky survey provides both deep multi-band photometry and homogeneous calibration across all MegaCam bands.

We therefore combine two external photometric catalogs to cover the necessary wavelength range. 
For the $g$, $r$, and $i$ bands, we adopt PS1~DR1, which reaches $5\sigma$ depths of 23.3, 23.2, and 23.1\,mag respectively over $\delta > -30^\circ$ \citep{chambers2016arXiv161205560C}. Since PS1 lacks a $u$-band, SDSS~DR12 is used as the $u$-band reference, providing a $5\sigma$ depth of approximately 22.0\,mag \citep{Alam_2015}. 

Because the MegaCam filter set differs from that of PS1 and SDSS, color-term corrections must be applied to place photometry in the MegaCam system. 
For clarity, the third-generation MegaCam filters installed in 2015 are labeled as \textit{u2:MP9302}, \textit{g2:MP9402}, \textit{r2:MP9602}, \textit{i3:MP9703}, and \textit{z2:MP9901}.

The converting relation from SDSS $u$-band filter to CFHT MegaCam $u2$-band filter is:

\begin{equation}
u2^{\rm CFHT} - u^{\rm SDSS} = -0.165 + 0.036 \times ( u^{\rm SDSS} - g^{\rm SDSS} )
\end{equation}

The converting relations from PS1 $g,r,i$-band filters to CFHT MegaCam $g2,r2,i3$-band filters are \footnote{https://www.cadc-ccda.hia-iha.nrc-cnrc.gc.ca/en/megapipe/docs/filt.html}:

\begin{multline}
g2^{\rm CFHT} - g^{\rm PS1} = 0.014 + 0.059 \times x - 0.00313 \times x^2 \\ 
- 0.00178 \times x^3,\ {\rm where}\ x=g^{\rm PS1} - i^{\rm PS1}; 
\end{multline}

\begin{multline}
r2^{\rm CFHT} - r^{\rm PS1} = 0.003 - 0.050 \times x + 0.01250 \times x^2  \\ 
- 0.00699 \times x^3,\ {\rm where}\ x=g^{\rm PS1} - i^{\rm PS1}; 
\end{multline}

\begin{multline}
i3^{\rm CFHT} - i^{\rm PS1} = 0.006 - 0.024 \times x + 0.00627 \times x^2   \\ 
- 0.00523 \times x^3,\ {\rm where}\ x=g^{\rm PS1} - i^{\rm PS1}; 
\end{multline}

As a reference, a full list of color terms is provided in Appendix~\ref{appendix:color_terms}. 
These coefficients are inputted into the configuration file of \texttt{processCcd.py} through the \texttt{calibrate.photoCal.colorterms} parameters. 

\begin{figure*}
    \centering
    \includegraphics[width=0.8\textwidth]{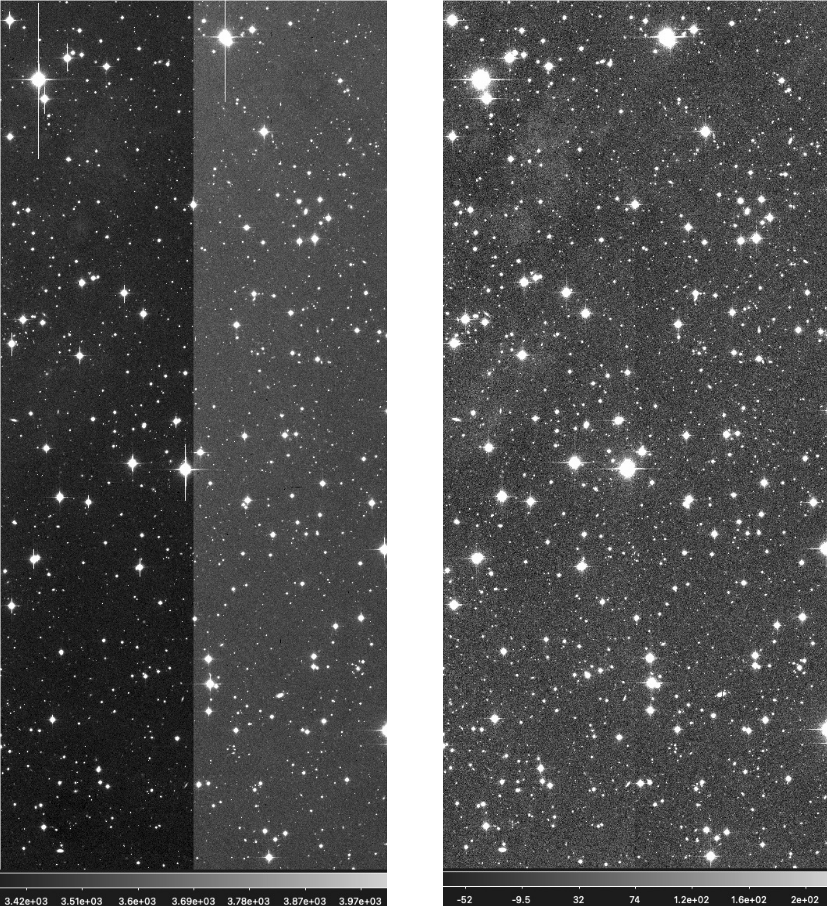}
    \caption{Single-frame processing result. \textit{left}: raw image of a single CFHT/MegaCam CCD in an exposure of the repeating FRB field; \textit{right}: the corresponding Processed Visit Image (PVI) from the 2DFIS pipeline.}
    \label{fig:single_frame_calexp_FRB}
\end{figure*}

After the step of calibration, Processed Visit Images (PVI) and Visit Source Catalogs (\texttt{src}) are generated for all single-epoch data. 
An example PVI is shown in Fig.~\ref{fig:single_frame_calexp_FRB}.
The PVI dataset comprises the processed image, which is detrended and background-subtracted, alongside an integer mask and a variance image offering per-pixel variance estimates. 
Each bit in the integer mask signifies distinct features like bad pixel, saturation or cosmic rays. 
The metadata and source catalogs include the astrometric calibration (WCS), the photometric calibration (including the magnitude zero-point), PSF solutions, aperture corrections, and sky background model. 
Regarding the specifics of source measurements, the catalogs include data on centroids, aperture and PSF photometry, adaptive-moment shapes, as well as preliminary morphological star-galaxy separation.

%%%%%%%%%%%%%%%%%%%%%%%%%%%%%%%%%%%%%%%%%%%%%%%%%%%%%%%%%%%%%%
%%%%%%%%%%%%%%%%%%%%%%%%%%%%%%%%%%%%%%%%%%%%%%%%%%%%%%%%%%%%%%
\subsection{Coaddition}
\label{subsec:coaddition}

After processing all single-frame exposures, we continue with the LSST Science Pipelines to coadd these exposures to generate a set of deep-coadded images.
In multi-epoch surveys, coadds are traditionally generated by reprojecting individual exposures onto a common sky grid and combining them to achieve greater depth. 
However, many standard coaddition approaches risk introducing systematic artifacts, such as correlated noise, PSF discontinuities, or astrometric distortions, which arise from repeated resampling or misaligned kernels.
The LSST Science Pipelines are designed to produce coadds that preserve the fidelity of the underlying imaging data while minimizing such systematics. 
The coaddition framework explicitly incorporates accurate WCS solutions, spatially varying PSF models, and robust background matching across CCDs, ensuring that depth is improved without compromising the photometric or morphological information essential for downstream analyses.

\subsubsection{Sky Map}

The coaddition process begins with the construction of a sky map that defines how the celestial sphere is partitioned for image reprojection and stacking. 
The LSST Science Pipelines adopt a hierarchical tiling scheme inspired by HEALPix \citep{Gorski_2005}, optimized for efficient data access and parallel processing. 
In this scheme, the sky is divided into large, rectangular regions called \textit{tracts}, which together provide a complete coverage of the survey footprint. 

Because each tract may span a substantial area on the sky and therefore require significant memory resources, it is further subdivided into smaller, regularly gridded \textit{patches}. 
These patches are defined in a local tangent-plane projection (TAN WCS), and neighboring patches overlap slightly to ensure continuity in source detection and measurement across boundaries. 
The subdivision is designed so that each patch fits comfortably within memory constraints, enabling efficient retrieval and independent processing.

For this work, we adopt a sky map composed of $5\times5$ patches per tract. 
Each patch spans $4000\times4000$ pixels at a pixel scale of $0\farcs185$ per pixel, and includes an additional 100-pixel border on all sides to provide overlap with adjacent patches. 
The calibrated single-epoch exposures (PVIs) are then reprojected onto these patches using the tract-level WCS, which locally approximates each patch as a planar region.

\subsubsection{Joint Calibration} 

Once the single-epoch images have been reprojected onto the sky map, it becomes possible to refine both astrometric and photometric calibrations by enforcing consistency in source positions and fluxes across all overlapping visits. 
This cross-visit information is essential for characterizing focal-plane distortions and for achieving uniform photometric calibration over the survey region.

The \texttt{jointcal} algorithm performs a simultaneous astrometric and photometric solution using all suitable stellar sources detected in the set of overlapping visits \citep{Bosch_2017}. 
It determines the model parameters by minimizing the residuals between the observed measurements and either (1) an external reference catalog or (2) internally fitted source parameters. 
The method relies on the expectation that isolated, non-variable point sources should exhibit identical celestial positions and fluxes across all visits. 
Incorporating multiple observations of the same stars significantly tightens the calibration constraints relative to those available from the reference catalogs alone. 

The astrometric model includes a 9th-order polynomial distortion function that delivers a smooth representation of the global focal-plane distortion. 
This distortion model is combined with per-CCD linear transformations (translation and rotation), which are assumed to remain constant across all visits within each tract-level fit. 
Additional visit-level polynomial terms capture time-dependent variations due to atmospheric refraction or telescope mechanical flexure.

For photometric calibration, \texttt{jointcal} anchors the overall solution to the external reference catalog, which is often shallower than the dataset being calibrated. 
The color terms derived in single-epoch processing are held fixed, and any remaining photometric adjustments are later refined on the coadd catalogs. 
By using all stellar detections, not only those in the reference catalog, \texttt{jointcal} effectively performs an internal calibration.
This improves spatial uniformity and provides finer corrections across the focal plane. 

The photometric model consists of a photometric zero-point and a 6th-order Chebyshev polynomial describing spatial variations across the focal plane for each visit. 
It also includes per-CCD scaling parameters that correct for residual flat-field errors and account for non-photometric observing conditions. 
Although these CCD-level terms are partially degenerate with atmospheric transparency variations, the multi-visit fitting helps to break these degeneracies and produce a stable photometric solution.

Further details on the implementation and mathematical formulation of \texttt{jointcal} can be found in \citet{Bosch_2017}.

\subsubsection{Make and Assemble Coadd}

The construction of coadded images begins with the \texttt{makeCoaddTempExp} stage, in which the calibrated single-epoch exposures (PVIs) are reprojected onto the predefined sky-map patches. 
During this resampling step, the astrometric and photometric solutions derived from \texttt{jointcal} are applied to ensure consistent alignment and flux calibration across visits. 
Depending on the configuration, the exposures may also undergo PSF matching, where each image is convolved with a kernel that homogenizes the PSF among visits. 
This optional PSF-matching step is useful for identifying transient features, such as asteroid trails or short-lived artifacts, by comparing PSF-homogenized exposures across epochs. 
% In our processing, these behaviors are controlled through the configuration parameters \texttt{doApplyUberCal=True} and \texttt{makePsfMatched=True}.

In addition to applying the \texttt{jointcal} calibrations, the pipeline optionally performs an internal relative photometric correction known as \texttt{UberCalibration}. 
Originally developed for the SDSS \citep{Padmanabhan_2008} and later adapted for other wide-field surveys (\textit{eg.} the DLS and PS1, \cite{Wittman_2012,Finkbeiner_2016}), \texttt{UberCalibration} uses a global least-squares optimization across overlapping exposures to correct spatial variations in photometric response. 
This procedure effectively accounts for residual flat-field errors and atmospheric transparency variations by utilizing the repeating observations in wide-field imaging. 
In our processing, we enable \texttt{doApplyUberCal=True} to improve large-scale photometric uniformity and reduce spatial systematics that are critical for color-based analyses and weak-lensing studies. 
% We also set \texttt{makePsfMatched=True} to produce PSF-matched images across bands, ensuring consistent photometric measurements for extended sources and enabling robust multi-band color measurements required for photometric redshift estimation.

The \texttt{assembleCoadd} stage then combines the individual resampled exposures into a final deep coadded image. 
The LSST Science Pipelines allow multiple types of temporary coadds to be produced, including both PSF-matched and direct (unmatched) coadds. 
It enables flexibility for different scientific applications. 
In our workflow, PSF-matched coadds are primarily used for artifact detection, since transient features stand out more clearly in exposures with homogenized PSFs. 
After identifying and flagging problematic regions or visits using these PSF-matched products, we construct our science-ready coadds using the direct exposures that remain after artifact rejection. 
This approach preserves the native PSF information and avoids unnecessary convolution, thereby retaining optimal image quality for downstream analysis such as shape measurement, photometry, and weak-lensing studies.

%%%%%%%%%%%%%%%%%%%%%%%%%%%%%%%%%%%%%%%%%%%%%%%%%%%%%%%%%%%%%%
%%%%%%%%%%%%%%%%%%%%%%%%%%%%%%%%%%%%%%%%%%%%%%%%%%%%%%%%%%%%%%
\subsection{Multi-band Processing}

Based on the deep coadded images, we carry out multi-band source detection, deblending, measurement, and forced photometry to produce the final object catalogs.

\subsubsection{Source Detection}

Source detection on the coadded images begins with the \texttt{detectCoaddSources} task. 
As in the single-epoch processing, each coadd is first smoothed with a matched filter and thresholded to identify significant peaks. 
Following the procedure used in the HSC-SSP pipeline \citep{Bosch_2017}, we adopt a nominal detection threshold of $5\sigma$ in PSF-flux signal-to-noise ratio. 
Thresholding produces contiguous regions of significant pixels, referred to as \texttt{footprints}, each of which records one or more local maxima. 
These peaks serve as provisional object centers and are subsequently refined during deblending.

Because coadded images reach substantially greater depth than single-epoch exposures, \texttt{footprints} frequently overlap or contain multiple astrophysical sources. 
After initial detection, each footprint is grown by an amount proportional to the PSF RMS to better capture the full spatial extent of the object. 
If this growth causes adjacent footprints to touch, they are merged to form larger, simply connected footprints that avoid double-counting while still preserving the peak structure needed for later deblending.

Multi-band detection is then handled by the \texttt{mergeCoaddDetection} task, which combines the detection catalogs from all coadded bands into a single master detection list. 
This merged catalog ensures that sources detected in any filter are included in the subsequent multi-band processing and that consistent apertures and deblending structures are used across all bands. 
The peaks from the different filters help to guide the separation of blended systems, reducing degeneracies in the identification of individual \textit{child} sources. 
Spurious peaks, particularly those caused by PSF wings or noise fluctuations around bright stars, are identified and removed at this stage to prevent contamination of the downstream measurements.

\subsubsection{Deblending}

The \texttt{deblendCoaddSources} step processes all detected \textit{parent} footprints that contain multiple significant peaks. 
Such footprints represent blends of two or more \textit{child} objects whose light profiles overlap in the coadded images. 
Deblending is performed using the \texttt{lsst.meas.deblender.SourceDeblendTask}, which is developed from the deblending algorithm for the \textit{SDSS Photo} pipeline \citep{Lupton_2001}. 
In this method, each detected peak is first compared to the local PSF model: peaks that match the PSF shape are treated as point-source templates, while extended peaks are modeled using symmetrized templates that capture non-stellar light distribution. 
The complete set of templates for all peaks in a footprint is then used to reconstruct the blended image, solving for non-negative template weights that best reproduce the observed pixel values. 
The resulting templates determine how flux is apportioned among the child objects, ensuring both non-negativity and exact conservation of the parent total flux. 

Compared to the traditional SExtractor deblender, the deblending algorithm of the LSST Science Pipelines offers several advantages. 
SExtractor relies on isophotal tree splitting and intensity cuts, which can misassign flux in crowded regions or produce unphysical shredding of bright galaxies. 
In contrast, the LSST deblender fits physically motivated templates constrained by the PSF and symmetry, leading to more stable separation of overlapping sources and greatly reducing the possibility to fragment extended galaxies. 
This template-fitting framework also yields fully flux-conserving and non-negative child images, which are essential for accurate photometry on deep coadds.

While the current LSST deblender operates in single-band mode, more advanced multi-band algorithms such as SCARLET \citep{Melchior_2018}, which uses joint spectral–morphological modeling to resolve degeneracies between blended objects, offer further improvements. 
We plan to explore the performance of SCARLET on our CFHT coadds in future work.

\subsubsection{Coadd Source Measurement}

Once deblending is completed, the \texttt{measureCoaddSources} task performs source measurements independently in each photometric band. 
On the deep coadds, the pipeline derives fundamental object properties, such as centroid positions, morphological parameters, and multiple flux estimates, using a suite of well-tested measurement algorithms. 
The effective PSF at each location on the coadd is constructed by combining the PSF models from the contributing single-epoch PVIs, ensuring that PSF estimates accurately reflect the varying seeing conditions of the input visits. 
Although measurements are performed separately in each band, the catalogs retain matched object identifiers across filters, which enables a direct band-to-band comparison. 

To create a unified multi-band catalog, the \texttt{mergeCoaddMeasurements} task consolidates the per-band measurement tables. 
For each object, the pipeline selects a \textit{reference band}, typically the one with the highest SNR and most reliable shape measurement. 
The reference band centroid and shape parameters are then adopted as the standard definitions of the morphology for all subsequent multi-band analyses. 
This approach minimizes inconsistencies that can arise when measurements are independently optimized in each band, particularly for faint or blended sources. 

Using these centroid positions and shape measurements, the \texttt{forcedPhotCoadd} task performs forced photometry on the coadded images in all filters. 
The fluxes are remeasured while all positional and morphological parameters are held fixed. 
This produces consistent multi-band photometry tied to a common aperture and morphology, which obtains accurate color measurements, well-defined spectral energy distributions (SEDs), and robust comparisons across filters even when an object is undetected or very faint in some bands.

A key advantage of the multi-band pipeline design is the strict separation between detection, shape estimation, and forced photometry. 
By anchoring the morphology to the band with the best data quality and enforcing the same model across all bands, the pipeline avoids band-to-band centroid misalignment, minimizes deblender inconsistencies, and suppresses color biases caused by noise. 
This strategy represents a major improvement over traditional per-band photometry ({e.g. SExtractor-style catalogs), providing significantly more stable and physically meaningful color measurements for faint galaxies in deep surveys.

\section{Data Products and Results}
\label{sec:data}

After processing by the LSST Science Pipelines, the final outputs consist of coadded images and object catalogs for each survey region. In this section, we present the 2DFIS data products, quality control measurements, and early scientific results.

\subsection{Data Products}
\label{subsec:data_products}

The data products of each deep field include image products and catalog products. 
As shown in Table \ref{tab:data_products}, both single-frame images (\texttt{calexp}) and single-band coadded images (\texttt{deepCoadd\_calexp}) are generated, along with multi-band color coadded images and weak lensing signal maps. 
In terms of catalog products, we generate and maintain single-frame source catalogs (\texttt{src}), single-band coadded catalogs (\texttt{deepCoadd\_src}), and merged multi-band object catalogs (combined \texttt{deepCoadd\_forced\_src}).

The raw and pre-processed image data are available on the Canadian Astronomy Data Centre (CADC \footnote{https://www.cadc-ccda.hia-iha.nrc-cnrc.gc.ca/en/}) website, while coadded image products and catalog products can be obtained upon request from the corresponding author.

\begin{table*}[ht]
    \centering
    \begin{tabular*}{\textwidth}{c c c c}
        \hline \hline
        \textbf{Processing Stage} & \textbf{Data Product} & \textbf{Pipeline Task} & \textbf{Description} \\ 
        \hline
        \multirow{4}{4em}{Single-frame processing} & calexp & processCcd & Processed visit images \\
        & calexpBackground & processCcd & Sky background in visit images \\
        & src   & processCcd & Single-frame source catalogs \\
        & icSrc & processCcd & Bright sources for initial CCD calibration \\
        \hline
        \multirow{2}{4em}{Joint calibration} & calib\_src & jointCal & UberCal source catalogs \\
        & calib\_exp & jointCal & UberCal exposures\\
        \hline
        \multirow{4}{4em}{Coadd processing} & deepCoadd\_calexp & measureCoaddSources & Coadded patch exposures \\
        & deepCoadd\_calexpBackground & measureCoaddSources & Sky background in coadded images \\
        & deepCoadd\_src & measureCoaddSources & Coadd source catalogs \\
        & deepCoadd\_forced\_src & forcedPhotCoadd & Forced photometry source catalogs \\
        \hline \hline
    \end{tabular*}
    \caption{Pipeline data products from each step}
    \label{tab:data_products}
\end{table*}

% Table: mask planes and catalog flags

% Table: source catalog entries

The multi-band color coadded image for the region of RXCJ0110.0+1358 (R.A.=01:10:00, Dec.=+13:58:22) is shown in Fig. \ref{fig:Cluster_coadd_vs_SDSS}. 
In comparison to the SDSS color image taken at the same pointing, the 2DFIS coadded image is significantly deeper and reveals more objects within the observed sky area.

\begin{figure*}
    \centering
    \includegraphics[width=1.0\textwidth]{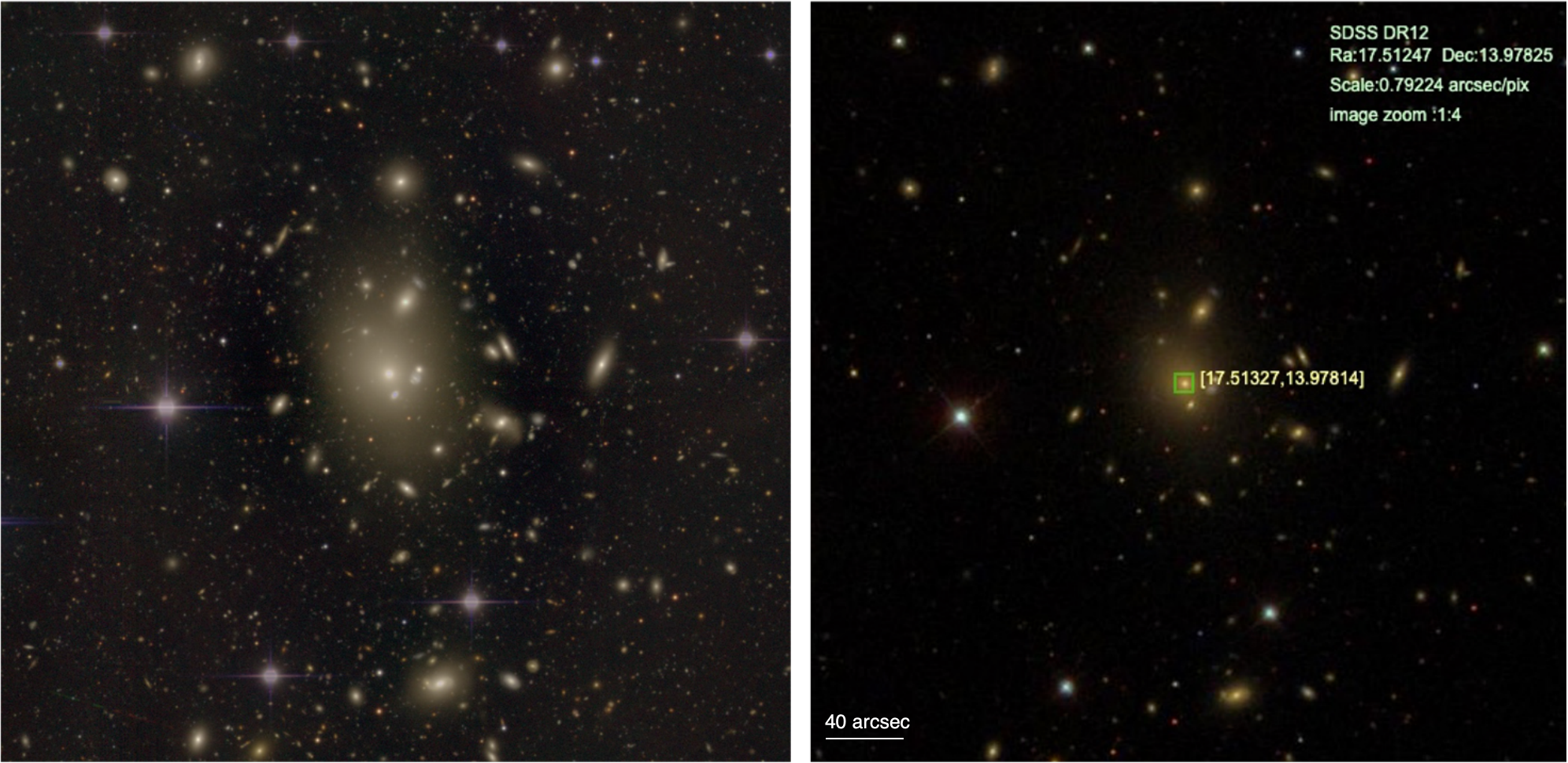}
    \caption{2DFIS coadd image result. \textit{left}: coadd image of the rotating cluster, composited from i/r/g-band; \textit{right}: SDSS image at the same sky area.}
    \label{fig:Cluster_coadd_vs_SDSS}
\end{figure*}

The color coadded image for the region of FRB190417 (R.A.=19:39:00, Dec.=+59:24:19) is shown in
Fig.~\ref{fig:FRB_coadd}, where the CHIME detection map is overlaid on top. 
The CHIME detection area ranges in R.A. 19:26:00 to 19:52:00, and Dec. +59:08:00 to +59:40:00.

\begin{figure*}
    \centering
    \includegraphics[width=0.8\textwidth]{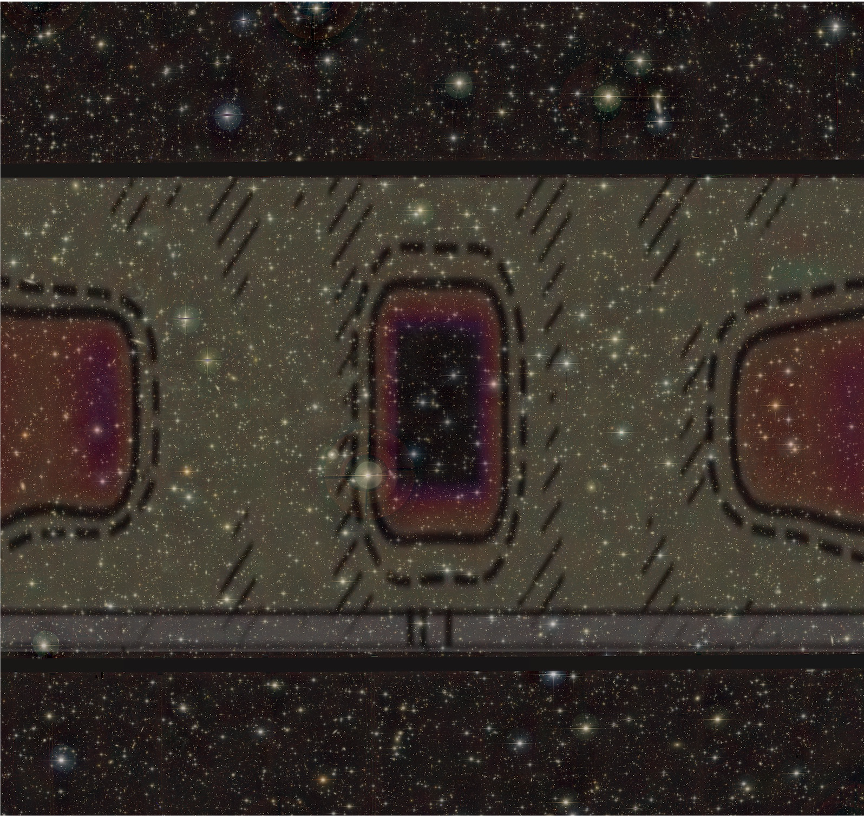}
    \caption{$1~\mathrm{deg}^2$-patch coadded image for the FRB repeater region, composited from i/r/g-band, overlapped with the CHIME detection area for the corresponding FRB events. The central dark region indicates the most possible sky area hosting the FRB source.}
    \label{fig:FRB_coadd}
\end{figure*}

Within the FRB repeater field, we identified a highly distorted galaxy at approximately $(19^{\mathrm h}38^{\mathrm m}26^{\mathrm s}, +59^\circ25'10'')$, located adjacent to a bright elliptical galaxy (Fig.~\ref{fig:FRB_coadd_arc_zoomin}).
The distorted source exhibits three brightness peaks and arc-like morphology, consistent with the visual appearance of a strong-lensing candidate.
Given that FRB190417 has three signal detections by CHIME \citep{CHIME_2019}, we initially considered the possibility that the distorted system might represent a lensed background host galaxy associated with the repeating FRB.
Photometric redshifts derived from the $u/g/r/i$ data placed the distorted galaxy behind the neighboring elliptical system, reinforcing the lensing hypothesis and motivating follow-up spectroscopy.
We therefore obtained spectroscopic observations with the 200-inch Hale Telescope (P200) to measure the redshifts of both galaxies.
The resulting spectra revealed that the two objects lie at comparable redshifts \citep{Wang_2025}, indicating that they form an interacting or merging pair rather than a lens–source configuration.
Although this rules out strong lensing in this case, the discovery highlights a dynamically disturbed system within the FRB field and demonstrates that optical counterparts or host candidates may be identifiable in this region.
A detailed analysis of this system will be presented in \cite{Wang_2025}.

\begin{figure*}
    \centering
    \includegraphics[width=1.0\textwidth]{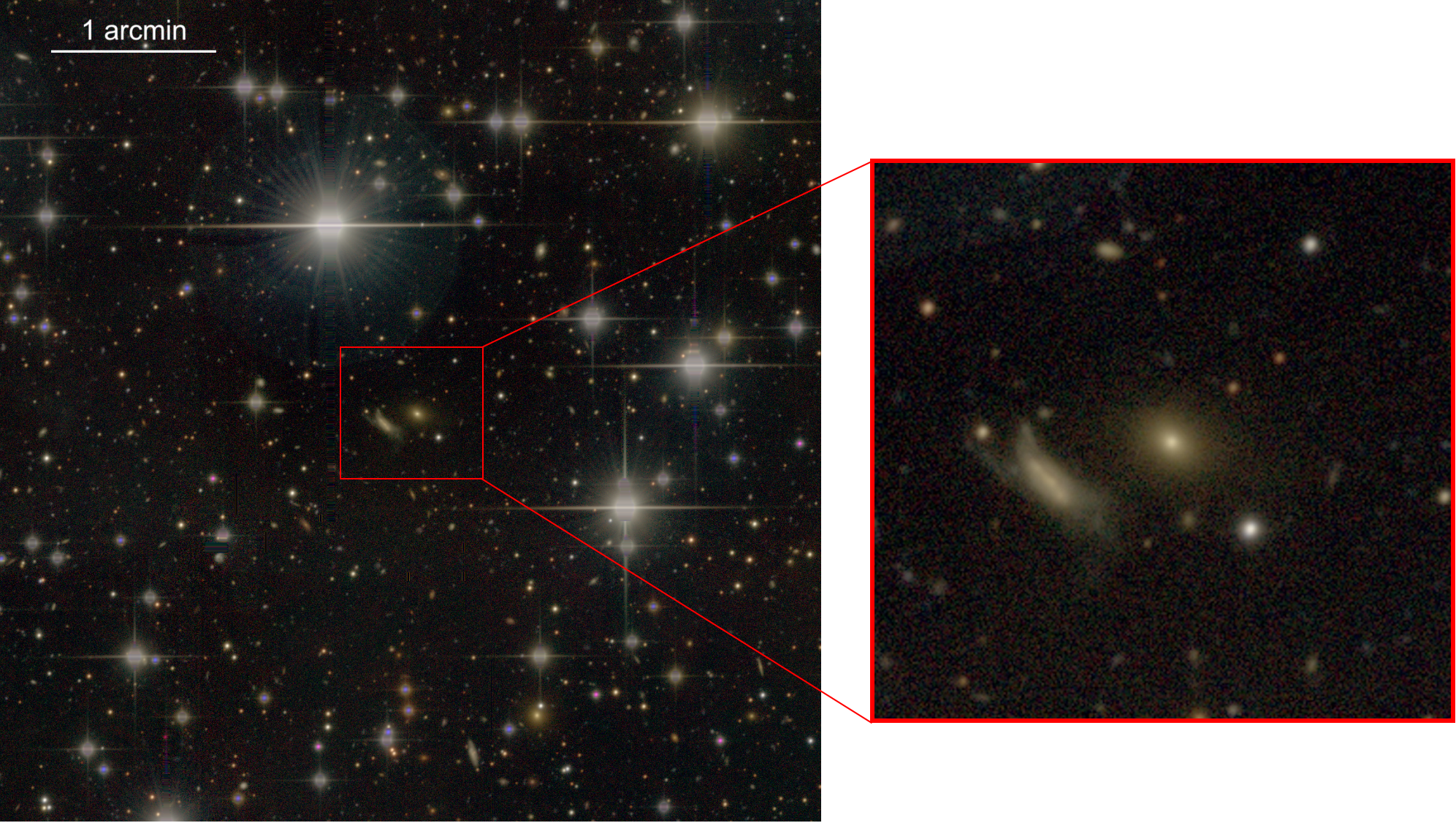}
    \caption{The i/r/g-band coadd image and a zoom-in for the suspect strong lensing candidate.}
    \label{fig:FRB_coadd_arc_zoomin}
\end{figure*}

\subsection{Quality Control (QC)}
\label{subsec:qc}

Before constructing coadded images, we perform a series of quality control (QC) checks on all single-epoch exposures to ensure that the PSF properties meet the requirements for weak-lensing and photometric analyses. 

In the LSST Science Pipelines, each single-epoch PVI source catalog includes the flag \texttt{calib\_psf\_used}, which identifies objects selected as PSF stars. 
Sources with \texttt{calib\_psf\_used = 1} are point-like, have adequate signal-to-noise, and are well described by the per-exposure PSF model.  
For every exposure, we extract this PSF-star sample and fit a two-dimensional Moffat profile at each star’s position \citep{Moffat_1969}. 
The FWHM values from these fits are used to characterize the effective PSF size for that exposure. 

To quantify PSF shape, we compute the second-moment ellipticity for each exposure using the adaptive second moments provided in the single-epoch catalogs. \footnote{In the LSST Science Pipelines v19.0, these second moments are produced by the \texttt{SdssShape} algorithm, which computes adaptive elliptical-Gaussian moments following the implementation used in SDSS imaging surveys.  
The algorithm iteratively adjusts the Gaussian weight function to match the object’s ellipticity and scale, producing robust measurements of $(I_{xx}, I_{yy}, I_{xy})$ even at modest signal-to-noise ratios.  
Although the resulting moments are technically weighted, they are widely adopted within the LSST pipeline as proxies for unweighted second moments and provide stable estimates of PSF size and ellipticity.}  
From the measured second moments $(I_{xx}, I_{yy}, I_{xy})$, we form: 
\begin{equation}
e_1 = \frac{I_{xx} - I_{yy}}{I_{xx} + I_{yy}}, \qquad
e_2 = \frac{2I_{xy}}{I_{xx} + I_{yy}},
\end{equation}
and define the second-moment ellipticity as
\begin{equation}
\epsilon = \sqrt{e_1^{\,2} + e_2^{\,2}}.
\end{equation}
% The same second moments also yield an estimate of the characteristic PSF size through the adaptive-moment radius
% \begin{equation}
% R_{\rm PSF} = \sqrt{I_{xx} + I_{yy}},
% \end{equation}
% which represents the width of the best-fitting adaptive Gaussian model.

Following the criteria established by the Local Volume Complete Cluster Survey (LoVoCCS; \citealt{Fu2022ApJ}), exposures used for weak-lensing shape measurements are required to have second-moment ellipticities $\epsilon < 0.14$ in the $r$ band and seeing better than $1''$ in the lensing band.
For photometric measurements in the other filters, LoVoCCS adopts a more permissive requirement of seeing $<1\farcs5$, which we also apply in the 2DFIS quality assessment. 
Representative PSF size and ellipticity distributions are shown in Fig.~\ref{fig:qc_cluster_2786625} and Fig.~\ref{fig:qc_frb_2785490} (more PSF QC plots are provided in Appendix~\ref{appendix:psf_qc}. 
All 2DFIS exposures satisfy these quality requirements.  
The rotating cluster field (22BD07) exhibits a median PSF FWHM below $0\farcs70$, while the FRB repeater field (22BS10) achieves a median FWHM of approximately $0\farcs50$, both well within the limits for weak-lensing analyses.
In both fields, the second-moment ellipticities remain comfortably below the $0.14$ threshold.

Given that all exposures meet the criteria for PSF size and shape, we retain the full dataset for subsequent coaddition and multi-band source measurement.

\begin{figure*}
    \centering
    \includegraphics[width=1.0\textwidth]{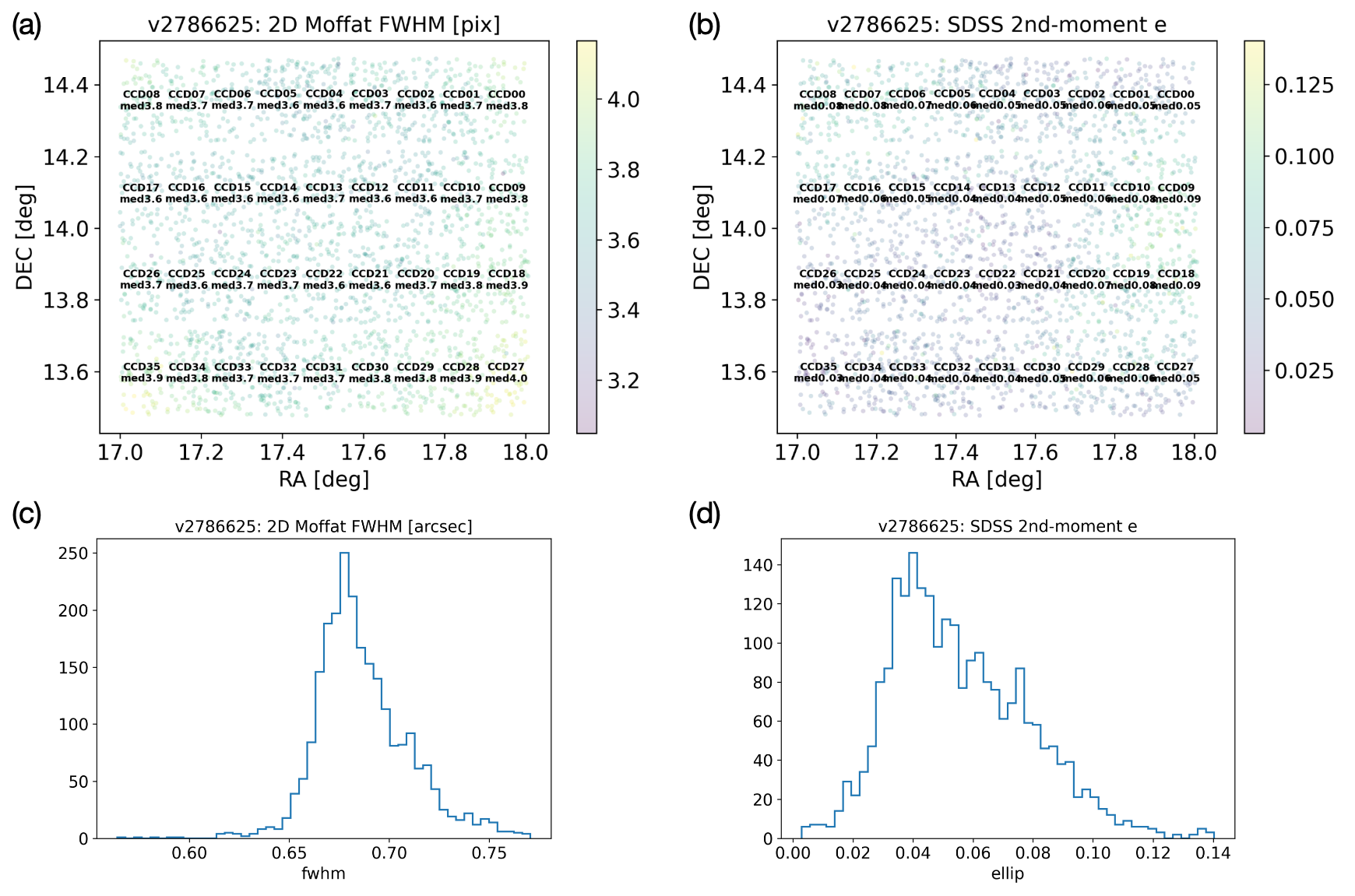}
    \caption{PSF QC for a single visit (2786625) in the rotating cluster field. (a) spatial distribution of the PSF star size in the focal plane, calculated as the Moffat function FWHM in pixel; (b) spatial distribution of the 2nd-moment ellipticity across the focal plane for the PSF stars; (c) histogram statistics of the Moffat function FWHM in arcsec; (d) histogram statistics of the 2nd-moment ellipticity.}
    \label{fig:qc_cluster_2786625}
\end{figure*}

\begin{figure*}
    \centering
    \includegraphics[width=1.0\textwidth]{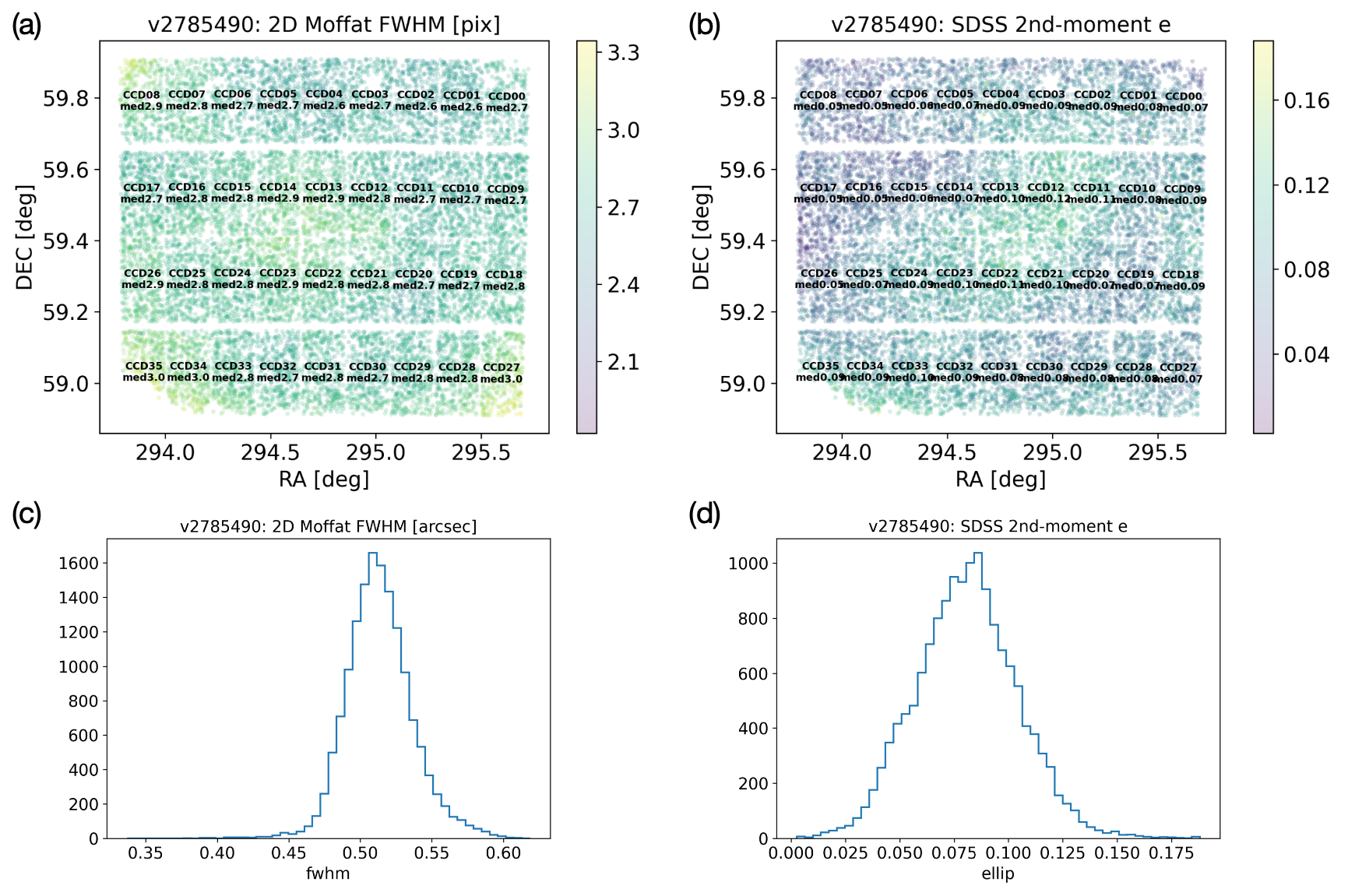}
    \caption{PSF QC for a single visit (2785490) in the FRB repeater field. (a) spatial distribution of the PSF star size in the focal plane, calculated as the Moffat function FWHM in pixel; (b) spatial distribution of the 2nd-moment ellipticity across the focal plane for the PSF stars; (c) histogram statistics of the Moffat function FWHM in arcsec; (d) histogram statistics of the 2nd-moment ellipticity.}
    \label{fig:qc_frb_2785490}
\end{figure*}

\subsection{Limiting Magnitude}
\label{subsec:limiting_mag}

As outlined in the observing proposals, the expected 5$\sigma$ depth of the 2DFIS data is approximately $r \sim 26.0$ for the rotating cluster field and $i \sim 25.2$ for the repeating FRB field.
To evaluate the achieved depths, we extract the fluxes and flux uncertainties from the coadded-source catalogs produced by the LSST pipelines.
For point sources we adopt the PSF flux, while for extended sources we use the \texttt{CModel} flux as an estimate of the total flux.
We then convert these fluxes and uncertainties to magnitudes and magnitude errors, and present the magnitude–error locus in Fig.\ref{fig:limiting_mag_cluster} and Fig.\ref{fig:limiting_mag_frb}, which provides a visual assessment of the survey depth.

\begin{figure*}
    \centering
    \includegraphics[width=0.8\textwidth]{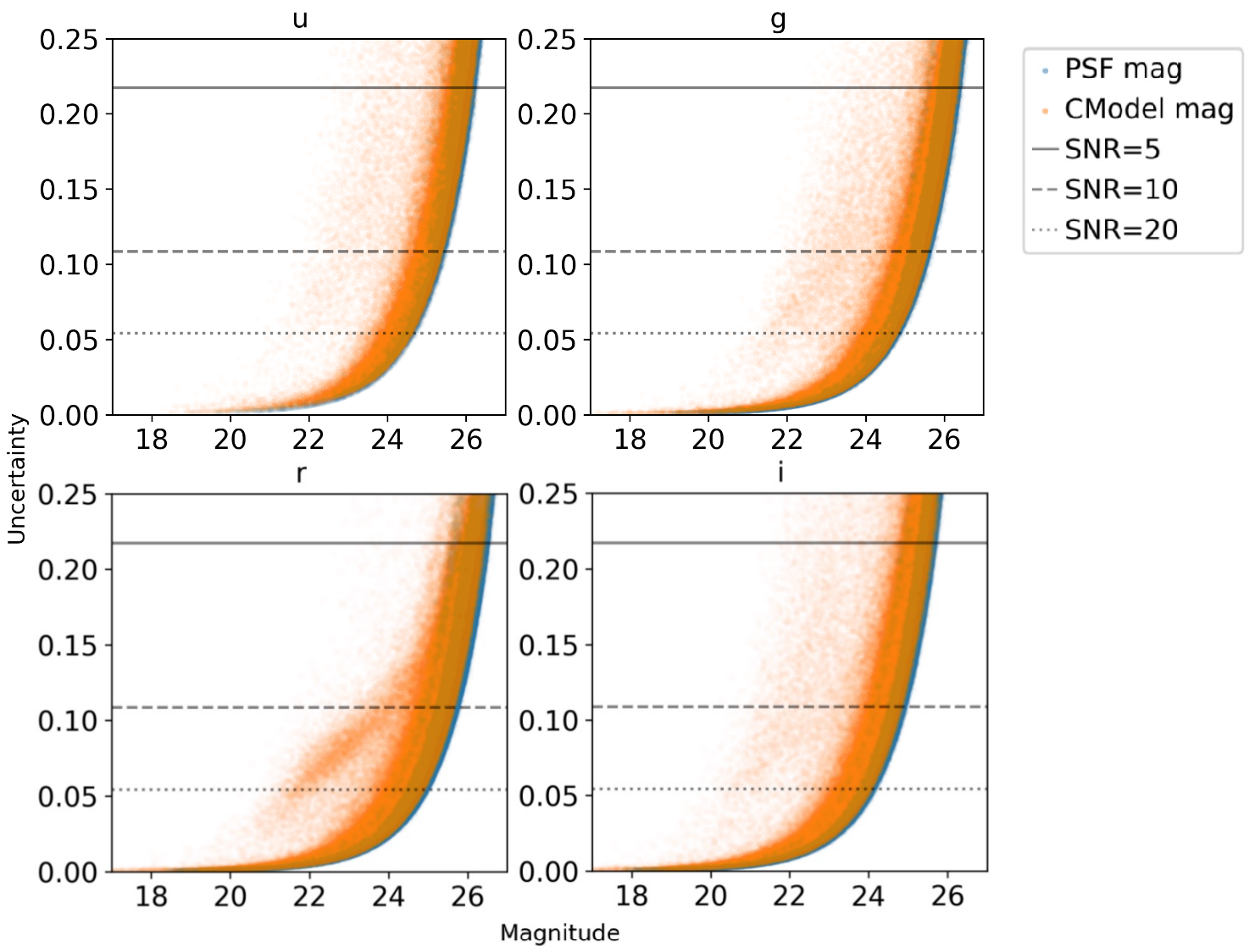}
    \caption{Source magnitude uncertainty vs. magnitude for the RXCJ0110.0+1358 field. Each panel is corresponding to a band. The point sources are denoted in blue dots, and the extended sources are in orange. A small number of sources form secondary clumps away from the main locus; such features are commonly attributed to blending, CCD-gap artifacts, or shredded detections from bright and extended galaxies \citep{Fu2022ApJ}.}
    \label{fig:limiting_mag_cluster}
\end{figure*}

\begin{figure*}
    \centering
    \includegraphics[width=0.8\textwidth]{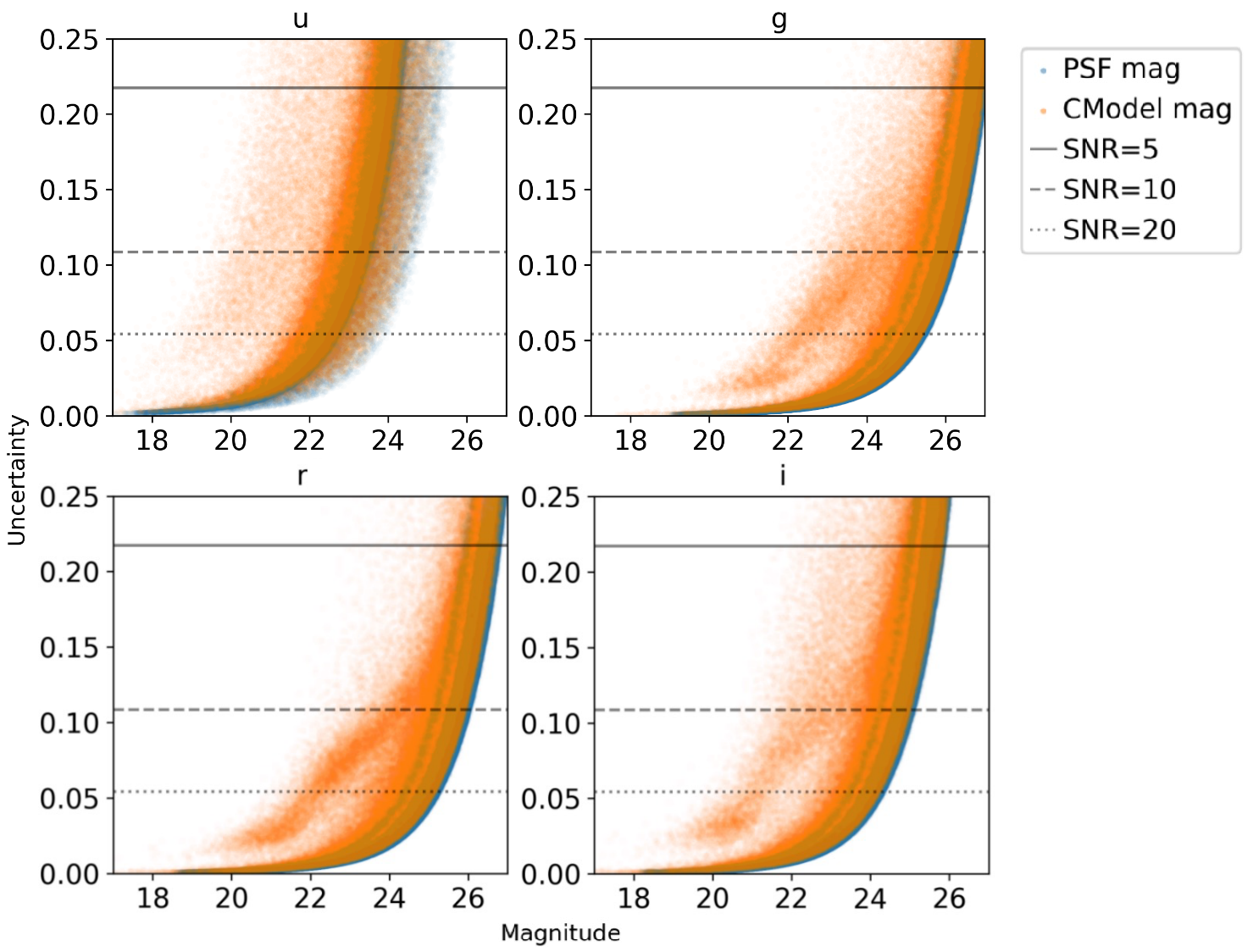}
    \caption{Source magnitude uncertainty vs. magnitude for the FRB190417 field. Each panel corresponds to a band. Point sources are shown as blue dots, and extended sources as orange dots. Similar secondary clumps offset from the main magnitude–error locus are present and are commonly attributed to blending, CCD-gap artifacts, or shredded detections from bright and extended galaxies. Owing to the deeper observations and higher source density in this field, source blending is expected to be more frequent.}
    \label{fig:limiting_mag_frb}
\end{figure*}

In addition, we compute the limiting magnitudes at different SNRs, which offers quantitative estimates of the observing depth in each field. 
For the rotating cluster field, the measured $r$-band depths at SNR = 10 reach $\mathrm{mag}_{\rm lim} = 25.78$ for extended sources and $25.95$ for point sources.
The remaining filters also achieve depths exceeding $25.0$mag, consistent with the expectations for this field.
For the FRB repeater field, the $r$-band limiting magnitudes at SNR = 10 are $26.09$ for extended sources and $26.18$ for point sources.
Across the other bands, the limiting depths span the range $24.76$–$26.38$mag, reflecting variation in exposure time and observing conditions.
A summary of the median magnitudes and limiting magnitudes for all filters is provided in Table.~\ref{tab:limiting_mags}.

The achieved depths in both fields are substantially deeper than those anticipated in the original proposals, reflecting the consistently good image quality throughout the observing runs.
With median seeing well below $1''$ and $r$-band limits approaching or exceeding $26$mag, the resulting coadds provide more than sufficient SNR for robust weak-lensing shear measurements, background-galaxy selection, and the photometric analyses required for both the cluster and FRB science goals.

\begin{deluxetable*}{llcccccccc}
\tablecaption{Limiting Magnitudes of the Two Fields\label{tab:limiting_mags}}
\tabletypesize{\small}
\tablehead{
\multicolumn{2}{c}{\textbf{RXCJ0110.0+1358 (22BD07)}} &
\multicolumn{2}{c}{$u$} &
\multicolumn{2}{c}{$g$} &
\multicolumn{2}{c}{$r$} &
\multicolumn{2}{c}{$i$} \\
\colhead{SNR} & \colhead{Sample} &
\colhead{median} & \colhead{max} &
\colhead{median} & \colhead{max} &
\colhead{median} & \colhead{max} &
\colhead{median} & \colhead{max}
}
\startdata
SNR=10 & star (PSF\_mag)        & 25.36 & 25.62 & 25.56 & 25.83 & 25.67 & 25.95 & 24.85 & 25.13 \\
       & galaxy (CModel\_mag)   & 25.08 & 25.55 & 25.23 & 25.74 & 25.24 & 25.78 & 24.43 & 25.00 \\
SNR=20 & star (PSF\_mag)        & 24.59 & 24.78 & 24.80 & 24.98 & 24.91 & 25.06 & 24.08 & 24.25 \\
       & galaxy (CModel\_mag)   & 24.29 & 24.69 & 24.46 & 24.89 & 24.49 & 24.96 & 23.66 & 24.19 \\
\tableline\tableline
\multicolumn{2}{c}{\textbf{FRB190417 (22BS10)}} &
\multicolumn{2}{c}{$u$} &
\multicolumn{2}{c}{$g$} &
\multicolumn{2}{c}{$r$} &
\multicolumn{2}{c}{$i$} \\
SNR & Sample &
median & max &
median & max &
median & max &
median & max \\
\tableline
SNR=10 & star (PSF\_mag)        & 23.46 & 24.85 & 26.19 & 26.44 & 25.95 & 26.18 & 25.01 & 25.26 \\
       & galaxy (CModel\_mag)   & 23.04 & 24.76 & 25.70 & 26.38 & 25.42 & 26.09 & 24.44 & 25.18 \\
SNR=20 & star (PSF\_mag)        & 22.66 & 24.02 & 25.42 & 25.61 & 25.19 & 25.35 & 24.25 & 24.41 \\
       & galaxy (CModel\_mag)   & 22.26 & 23.84 & 24.93 & 25.49 & 24.66 & 25.24 & 23.70 & 24.29 \\
\enddata
\end{deluxetable*}

\section{Discussion}
\label{sec:discussion}

\subsection{Systematic Uncertainties}
\label{subsec:systematics}

The accuracy of weak-lensing and photometric measurements in 2DFIS depends critically on controlling both observational and data-processing systematics. 
Despite meeting the required image-quality specifications, CFHT MegaCam observations are subject to small variations in seeing, instrumental response, and photometric calibration that can lead to correlated uncertainties in shape and flux measurements.
Below, we summarize the dominant sources of systematic uncertainty associated with both the observing strategy and data reduction.

\subsubsection{Observational Systematics}

The primary observational uncertainty arises from variations in seeing and transparency across exposures. 
Even within the required $r$-band image quality of 0.65–0.80\arcsec, short-term fluctuations in the PSF profile can introduce spatially correlated shape distortions that mimic weak-lensing shear.  
To mitigate this, the exposures in each filter were overlapped with dithers at multiple sky positions, enabling a well-sampled PSF reconstruction and minimizing chip-edge systematics.  
Residual PSF anisotropy, particularly at the corners of the MegaCam focal plane, contributes a shape-measurement bias at the level of $\sim10^{-3}$ in ellipticity, consistent with previous CFHTLenS analyses \citep{Miller_2013}. 

Sky brightness variations also affect photometric precision.  
The cluster RXCJ0110.0+1358 field was observed under gray conditions, while the FRB190417 field was taken during darker time but at a higher airmass ($1.5 \lesssim X \lesssim 2.0$).  
This leads to small differences in atmospheric extinction and background level between filters, which are accounted for in the photometric calibration using standard star fields and global color terms. 
Differential atmospheric refraction across the field-of-view is another potential source of astrometric error, but its impact is limited to $\lesssim0.05''$ under the adopted observing constraints \citep{Lu_2025}.

\subsubsection{Data Reduction and Analysis Systematics}

All data were processed using the LSST Science Pipelines v19.0 \citep{Bosch_2019}, with an interface package adapted for CFHT MegaCam data. 
Several systematic effects inherent to the processing chain were carefully examined.

The astrometric calibration uncertainties arise from the choice of reference catalogs and from local distortions across the focal plane. 
Residual offsets between individual CCDs after joint astrometric fitting are typically below 30~mas, as verified against the {\it Gaia}~DR3 reference frame.  
Such subpixel-level errors are negligible for photometric measurements but contribute to the effective PSF modeling uncertainty in shape analyses.

The photometric calibration uncertainties stem from the determination of zero points and color terms. 
The LSST Science Pipelines derive per-exposure photometric solutions using standard stars and internal consistency across overlapping fields. 
For 2DFIS, the resulting zero-point stability is better than 0.02~mag across the field, which dominates the uncertainty in color-based stellar locus correction and photometric redshift estimation. 

The PSF modeling errors are among the most significant sources of systematic bias in weak-lensing applications.  
The pipeline constructs a spatially varying PSF model using shapelet or PCA-based interpolation of stellar profiles.  
Residuals between the modeled and observed PSFs typically remain below 1–2\% in size and 0.5\% in ellipticity.  
These residuals are further monitored and incorporated into the shear calibration uncertainty budget. 

% The background estimation and coaddition systematics influence faint object detection near the survey limit.  
% The global sky background in each exposure is modeled with a 6th-order Chebyshev polynomial fit in the LSST Science Pipelines, which may over-subtract extended low-surface-brightness (LSB) features. 
% To minimize this effect, another round of local background subtraction is performed during the image characterization step and we visually inspect the PVIs for artifacts. 
% The local background removal may introduce over-subtraction around bright objects, which can be noticed in Fig.~\ref{fig:Cluster_coadd_vs_SDSS}. 
% But it has negligible impact on the photometric profile of faint and small objects. 
% Given that the background sample in the mass map reconstruction is dominated by small objects, this global-local background subtraction has limited influence on the weak lensing analysis. 
% In the future, different parameters of background removal will be adopted depending on specific science cases. 
Background estimation systematics can affect the detection of faint sources near the survey limiting depth. 
In the LSST Science Pipelines, the global sky background for each single-epoch exposure is modeled using a 6th-order Chebyshev polynomial. 
While this approach effectively captures large-scale gradients, it can bias the extended low-surface-brightness (LSB) sources. 
To mitigate this, an additional local background subtraction is applied during the image characterization stage, and all PVIs are visually inspected for possible artifacts. 
Although the local subtraction can lead to slight oversubtraction in the vicinity of bright objects—visible in Fig.~\ref{fig:Cluster_coadd_vs_SDSS}, its impact on faint galaxies is minimal. 
Because the weak-lensing mass reconstruction relies predominantly on such small background sources, the combined global and local background removal has only a limited effect on our shear measurements. 
Future analyses may adopt background-removal configurations tailored to the specific scientific requirements of different fields.

Overall, the combination of stable observing conditions, dithered image coverage, and robust LSST pipeline processing ensures that the residual systematic uncertainties in 2DFIS are well below the statistical errors for both weak-lensing shape measurements and photometric analyses.  
Future reprocessing with more recent pipeline versions (e.g., v29.0 or later) and improved PSF modeling tools such as \texttt{PIFF} \footnote{\url{https://rmjarvis.github.io/Piff}} will further reduce these residual biases.

\subsection{Science Goals}
\label{subsec:science_goals}

The 2DFIS is designed to exploit the wide-field imaging capability of CFHT MegaCam and the robust analysis framework of the LSST Science Pipelines to address two complementary scientific objectives: 
(1) mapping the mass distribution of a dynamically complex galaxy cluster through weak gravitational lensing, and 
(2) characterizing the dark and luminous environment surrounding a repeating FRB source.  
The survey also provides a valuable dataset for a variety of secondary science applications involving galaxy evolution, large-scale structure, and time-domain phenomena.

\subsubsection{Mass Mapping of a Rotating Galaxy Cluster}

The first 2DFIS field targets a nearby galaxy cluster (RXCJ0110.0+1358, $z\simeq0.058$) exhibiting signatures of rotational or merging motion inferred from redshift asymmetries of its member galaxies.  
The deep multi-band imaging obtained in $u$, $g$, $r$, and $i$ allows accurate measurement of the weak-lensing shear field over the entire $1\,\deg^2$ MegaCam field of view.  
By combining these lensing-derived shear maps with available X-ray and spectroscopic data, we can reconstruct the projected mass distribution and test whether the observed velocity asymmetry originates from large-scale rotation, a line-of-sight merger, or multiple subcluster components.  
This analysis enables a quantitative comparison between the baryonic (X-ray emitting gas) and dark matter distributions, which provides new constraints on cluster dynamics and dark matter–baryon coupling in non-relaxed systems.  
The data will also serve as an independent test case for shear calibration and mass calibration methods applicable to next-generation wide-field surveys.

\subsubsection{The Dark Environment of a Repeating FRB Source}

The second field in 2DFIS targets the repeating FRB (FRB190417), chosen to probe the optical and dark-matter environment of a high-declination FRB field.  
With deep imaging down to $r\simeq26$~mag, the survey enables a systematic search for potential host-galaxy candidates and for faint optical counterparts within the FRB localization region. 
Weak-lensing analysis of the same field provides a direct measure of the projected matter density along the line of sight, which can be compared to that of other known FRB fields such as FRB190520B \citep{Chen_2025}. 
By combining optical imaging, environmental characterization, and lensing measurements, 2DFIS can examine if repeating FRBs tend to reside in overdense magnetized regions or in typical field-galaxy populations. 
This comparison provides clues as to whether repeating and non-repeating FRBs arise from a shared origin or from distinct progenitor mechanisms.

% \textcolor{red}{arc in this field: photo-z and spec-z to rule out strong lensing --> tidelly distorted galaxy }
% \textcolor{red}{zoom-in galaxies. Spectrum. We found these galaxies belong to one system.}
% -- Added in section 4.1

\subsubsection{Legacy and Ancillary Science}

Beyond the primary science goals, the 2DFIS dataset provides deep imaging in four bands with precise astrometric and photometric calibration, making it suitable for a broad range of ancillary investigations. 
The coadded catalogs enable studies of galaxy color–magnitude relations, star–galaxy separation, and photometric redshift calibration for faint sources.  
The large-area coverage and depth also allow the identification of strong-lensing features, LSB structures, and intracluster light (ICL) around the targeted cluster.  
For the FRB190417 field, time-domain cross-matching with existing radio and X-ray archives can be used to search for variable or transient counterparts. 
Moreover, the 2DFIS data products, including single-frame calibrated images, coadds, and source catalogs, provide a useful testbed for validating LSST-like reduction pipelines on non-LSST instruments, supporting future synergies among LSST, CFHT, and other upcoming wide-field surveys. 

In summary, 2DFIS serves its originally proposed scientific objectives: weak-lensing confirmation of a dynamically complex cluster and environmental mapping of a repeating FRB. It also creates a legacy dataset of broader utility for extragalactic, cosmological, and time-domain research. 
\section{Conclusions and Future Work}
\label{sec:conclusions}

In this study, we introduced the Two Deep Fields Imaging Survey (2DFIS), utilizing CFHT data to investigate two scientifically intriguing fields: one centered on a repeating FRB event field and the other on a dynamically ``rotating'' galaxy cluster. 
By employing the LSST Science Pipelines with custom adaptations for CFHT data, we successfully processed the imaging data to produce calibrated single-epoch exposures, high-quality multi-band coadded images, and detailed source catalogs. 
These data products enable a range of analyses, including the search for optical counterparts to the FRB and the study of mass distributions in the galaxy cluster via weak gravitational lensing. 
The subsequent weak lensing analysis, including shear calibration, photo-z estimation, and mass distribution, will be provided in \cite{Yicheng_2025}. 

While the pipeline adaptations yielded robust results, challenges were identified that provide improvements in future work. 
In particular, enhancing the multi-band data processing framework to better address discrepancies across filters will be a key focus. 
Refinements in photometric calibration and coaddition techniques will also improve the accuracy and consistency of multi-band products. 

Additionally, mitigating oversubtraction in crowded or extended source regions is critical for preserving the fidelity of faint source detections and shape measurements. 
Future efforts will focus on implementing advanced deblending algorithms and optimizing background subtraction to reduce these effects.

Ongoing improvements to the calibration and analysis pipelines will enhance the precision and consistency of 2DFIS data products, which will enable more robust weak-lensing measurements, photo-z estimates, and transient detections. 
Future survey extensions will target additional deep fields to increase the statistical power for cluster mass calibration and to sample a wider range of FRB host environments.
We also plan to integrate spectroscopic observations and multi-wavelength data, and further allow direct comparisons between baryonic and dark matter tracers. 
Through iterative refinement of the current processing workflow and the adoption of advanced techniques such as GPU-accelerated image analysis and machine-learning classification, 2DFIS will continue to provide high-quality datasets for scientific cases in both static and time-domain cosmology.

\begin{acknowledgements}

\textit{Acknowledgements:} This work is based on data obtained as part of the Canada-France Imaging Survey, a CFHT large program of the National Research Council of Canada and the French Centre National de la Recherche Scientifique. Based on observations obtained with MegaCam, a joint project of CFHT and CEA Saclay, at the Canada-France-Hawaii Telescope (CFHT) which is operated by the National Research Council (NRC) of Canada, the Institut National des Science de l'Univers (INSU) of the Centre National de la Recherche Scientifique (CNRS) of France, and the University of Hawaii.

We would like to thank the Canada-France-Hawaii Telescope (CFHT) Operations, Instrumentation, and Software Groups for their contributions and diligence in maintaining observatory operations; the CFHT Astronomy Group for their observation coordination and data acquisition efforts; and the CFHT Finance \& Administration Group for their contributions to the management and administration of the observatory.

BL acknowledges the support from the National Key Research and Development Program of China (2023YFA1608100), and the Double-Innovation Doctor Program of Jiangsu Province (No. JSSCBS0216). 
WL acknowledges the support from the National Key R\&D Program of China (2021YFC2203100, 2018YFA0404503), NSFC(NO. 12573005, 11833005, 12192224, 11890693), CAS Project for Young Scientists in Basic Research, Grant No. YSBR-062, the Fundamental Research Funds for the Central Universities, and the China Manned Space Project with No. CMS-CSST-2021-A03. 

This research was conducted using computational resources and services at the Center for Computation and Visualization (CCV) of Brown University, and the Starburst Computing Platform of Purple Mountain Observatory. 

This research utilizes software created for the Legacy Survey of Space and Time conducted by the Vera C. Rubin Observatory. Our gratitude goes to the LSST Project for providing their code as open-source, accessible at \url{http://dm.lsstcorp.org}. 

This work has made use of data from the European Space Agency (ESA) mission {\it Gaia} (\url{https://www.cosmos.esa.int/gaia}), processed by the {\it Gaia} Data Processing and Analysis Consortium (DPAC, \url{https://www.cosmos.esa.int/web/gaia/dpac/consortium}). Funding for the DPAC has been provided by national institutions, in particular the institutions participating in the {\it Gaia} Multilateral Agreement.

The Pan-STARRS1 Surveys (PS1) and the PS1 public science archive have been made possible through contributions by the Institute for Astronomy, the University of Hawaii, the Pan-STARRS Project Office, the Max-Planck Society and its participating institutes, the Max Planck Institute for Astronomy, Heidelberg and the Max Planck Institute for Extraterrestrial Physics, Garching, The Johns Hopkins University, Durham University, the University of Edinburgh, the Queen's University Belfast, the Harvard-Smithsonian Center for Astrophysics, the Las Cumbres Observatory Global Telescope Network Incorporated, the National Central University of Taiwan, the Space Telescope Science Institute, the National Aeronautics and Space Administration under Grant No. NNX08AR22G issued through the Planetary Science Division of the NASA Science Mission Directorate, the National Science Foundation Grant No. AST-1238877, the University of Maryland, Eotvos Lorand University (ELTE), the Los Alamos National Laboratory, and the Gordon and Betty Moore Foundation.

Funding for the Sloan Digital Sky Survey V has been provided by the Alfred P. Sloan Foundation, the Heising-Simons Foundation, the National Science Foundation, and the Participating Institutions. SDSS acknowledges support and resources from the Center for High-Performance Computing at the University of Utah. 
SDSS telescopes are located at Apache Point Observatory, funded by the Astrophysical Research Consortium and operated by New Mexico State University, and at Las Campanas Observatory, operated by the Carnegie Institution for Science. The SDSS web site is \url{www.sdss.org}.

SDSS is managed by the Astrophysical Research Consortium for the Participating Institutions of the SDSS Collaboration, including the Carnegie Institution for Science, Chilean National Time Allocation Committee (CNTAC) ratified researchers, Caltech, the Gotham Participation Group, Harvard University, Heidelberg University, The Flatiron Institute, The Johns Hopkins University, L'Ecole polytechnique f\'{e}d\'{e}rale de Lausanne (EPFL), Leibniz-Institut f\"{u}r Astrophysik Potsdam (AIP), Max-Planck-Institut f\"{u}r Astronomie (MPIA Heidelberg), Max-Planck-Institut f\"{u}r Extraterrestrische Physik (MPE), Nanjing University, National Astronomical Observatories of China (NAOC), New Mexico State University, The Ohio State University, Pennsylvania State University, Smithsonian Astrophysical Observatory, Space Telescope Science Institute (STScI), the Stellar Astrophysics Participation Group, Universidad Nacional Aut\'{o}noma de M\'{e}xico, University of Arizona, University of Colorado Boulder, University of Illinois at Urbana-Champaign, University of Toronto, University of Utah, University of Virginia, Yale University, and Yunnan University.

\end{acknowledgements}

\software{
The LSST Science Pipelines \citep{Bosch_2019}, 
astropy \citep{2013A&A...558A..33A,2018AJ....156..123A}
}

%\beforebibliography
\bibliographystyle{aasjournal}
\bibliography{biblio}

\appendix

\section{CFHT--PS1 Photometric Color Transformations}
\label{appendix:color_terms}

Accurate photometric calibration of CFHT/MegaCam data requires transforming instrumental magnitudes onto a stable external reference system.  
For 2DFIS, we adopt the Pan-STARRS1 (PS1) photometric system as the absolute reference and apply the color-term corrections defined for MegaCam when calibrating $g$, $r$, $i$, $z$ bands against PS1 stellar photometry.  
These relations account for differences in effective throughput between the CFHT and PS1 filter sets and are required to place CFHT magnitudes on a uniform AB system.  
CFHT/MegaCam filters have undergone several upgrades since commissioning, resulting in three distinct generations of filters.\footnote{A detailed description of the MegaCam filter generations, including transmission curves and installation history, is available at: \url{https://www.cadc-ccda.hia-iha.nrc-cnrc.gc.ca/en/megapipe/docs/filt.html}.}

All color terms are expressed as a function of the PS1 color
\begin{equation}
x = g^{\rm PS1} - i^{\rm PS1},
\end{equation}
and take the polynomial form
\begin{equation}
m^{\rm CFHT} - m^{\rm PS1}
    = a_0 + a_1 x + a_2 x^2 + a_3 x^3,
\end{equation}
where the coefficients $(a_0, a_1, a_2, a_3)$ depend on the specific MegaCam filter.

Below we list the transformations for both the first/second-generation MegaCam filters (installed prior to 2015) and the third-generation filters (installed in 2015 and being used at present).  
These equations follow the official CFHT calibration recommendations for PS1-anchored photometry.

\subsection{First and Second Generation MegaCam Filters (Pre-2015)}

The following color terms apply to the first- and second-generation filters: MP9401, MP9601, MP9701, MP9702, and MP9801.
\begin{align}
g({\rm MP9401}) - g^{\rm PS1} &= -0.001 - 0.004\,x - 0.0056\,x^2 + 0.00292\,x^3, \\
r({\rm MP9601}) - r^{\rm PS1} &= 0.002 - 0.017\,x + 0.00554\,x^2 - 0.000692\,x^3, \\
i({\rm MP9701}) - i^{\rm PS1} &= 0.001 - 0.021\,x + 0.00398\,x^2 - 0.00369\,x^3, \\
i2({\rm MP9702}) - i^{\rm PS1} &= -0.005 + 0.004\,x + 0.0124\,x^2 - 0.0048\,x^3, \\
z({\rm MP9801}) - z^{\rm PS1} &= -0.009 - 0.029\,x + 0.012\,x^2 - 0.00367\,x^3.
\end{align}

\subsection{Third Generation MegaCam Filters (2015–Present)}

The third-generation filters MP9402, MP9602, MP9703, and MP9901 provide improved throughput and stability, and have been used for all CFHT imaging since 2015.  
Their corresponding color terms relative to PS1 are:
\begin{align}
g2({\rm MP9402}) - g^{\rm PS1} &= 0.014 + 0.059\,x - 0.00313\,x^2 - 0.00178\,x^3, \\
r2({\rm MP9602}) - r^{\rm PS1} &= 0.003 - 0.05\,x + 0.0125\,x^2 - 0.00699\,x^3, \\
i3({\rm MP9703}) - i^{\rm PS1} &= 0.006 - 0.024\,x + 0.00627\,x^2 - 0.00523\,x^3, \\
z2({\rm MP9901}) - z^{\rm PS1} &= -0.016 - 0.069\,x + 0.0239\,x^2 - 0.0056\,x^3.
\end{align}

\subsection{Application in the 2DFIS Calibration Pipeline}

During photometric calibration, each matched PS1 reference star is corrected using the appropriate color-term equation based on the MegaCam filter generation used for the exposure.  
The color-corrected PS1 magnitude is then compared with the instrumental CFHT magnitude to derive the photometric zero point.  
Applying these transformations ensures consistent calibration across all filters and suppresses color-dependent systematics in the final coadded catalogs.

These corrections are applied automatically during the \texttt{photoCalib} stage of the LSST Science Pipelines, ensuring that all 2DFIS photometry is placed on a uniform PS1-based AB magnitude system.

\section{PSF Quality Control Plots}
\label{appendix:psf_qc}

We summarize the $r$-band PSF FWHM and ellipticity quality-control results for all exposures in both fields.
Visits 2786623, 2786624, 2786625, and 2786626 correspond to the rotating cluster field (22BD07), while visits 2785487, 2785488, 2785489, and 2785490 correspond to the repeating FRB field (22BS10).
For each set of visits, panel (a) shows the spatial distribution of PSF FWHM across the field of view, and panel (b) shows the corresponding ellipticity distribution.
Panels (c) and (d) present the histograms of PSF FWHM and ellipticity, respectively.

\begin{figure*}
    \centering
    \includegraphics[width=0.9\textwidth]{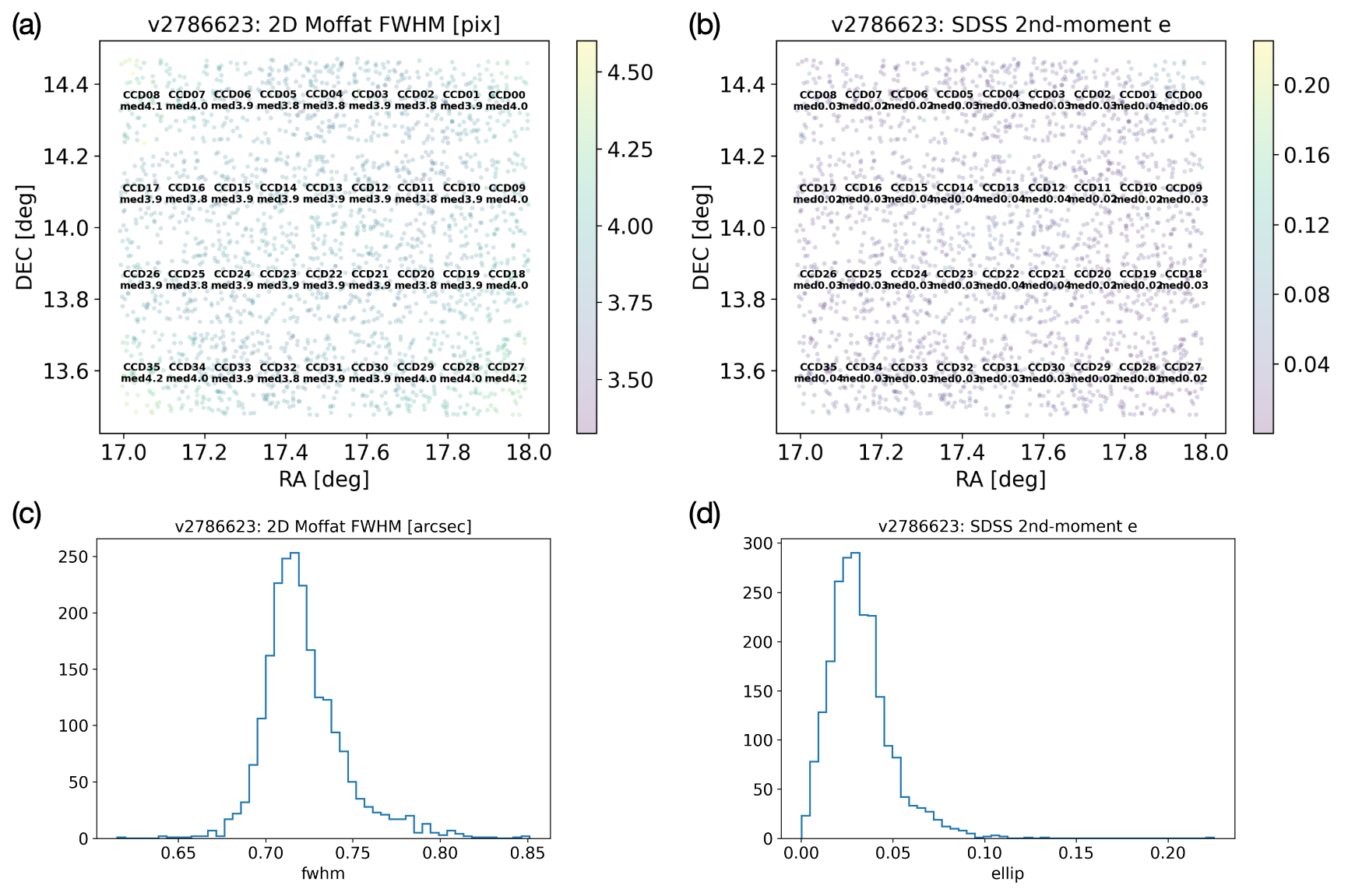}
    \includegraphics[width=0.9\textwidth]{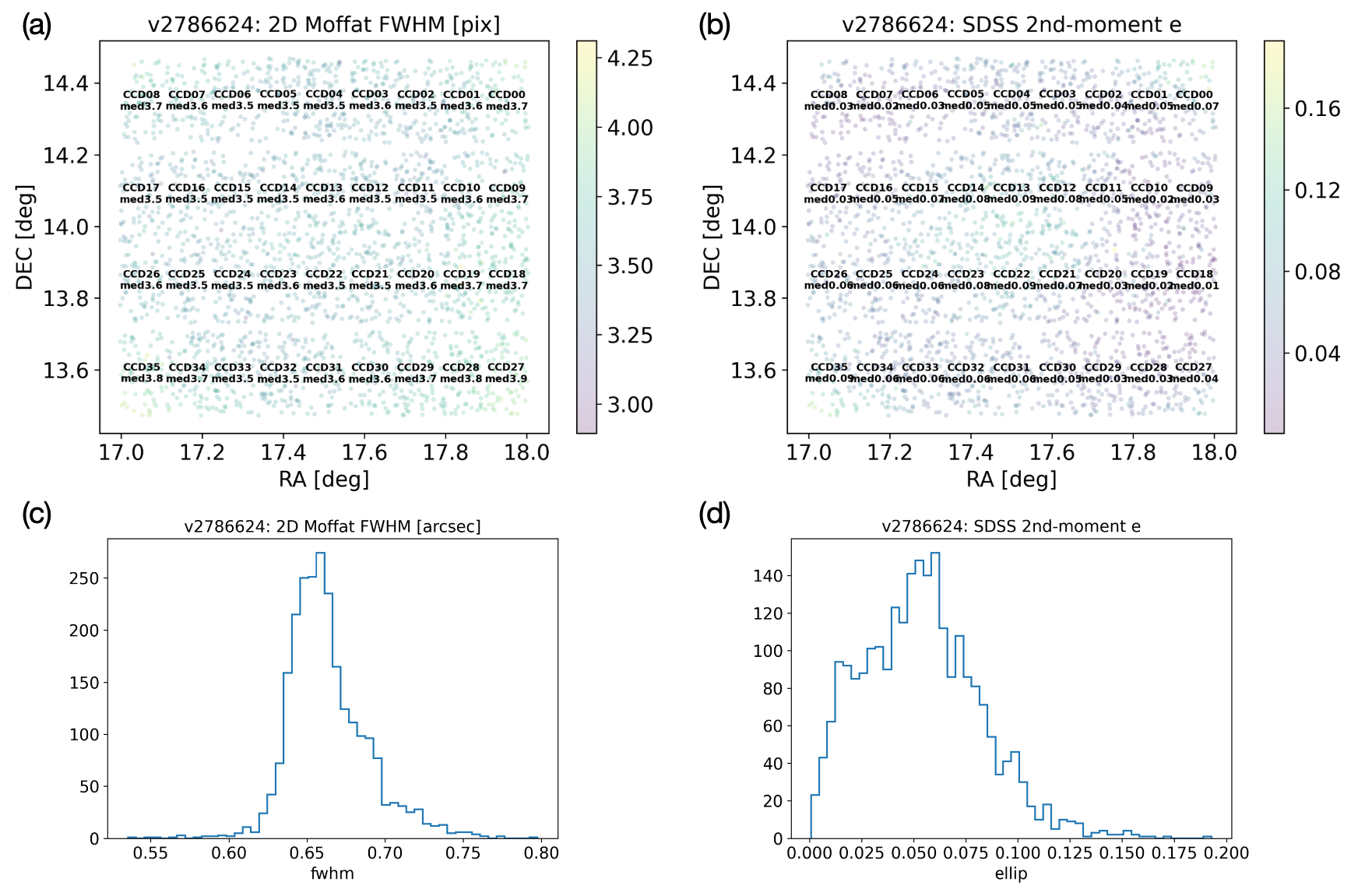}
    \caption{PSF QC for visit 2786623 and 2786624.}
    \label{fig:qc_cluster_2786623}
\end{figure*}

\begin{figure*}
    \centering
    \includegraphics[width=0.9\textwidth]{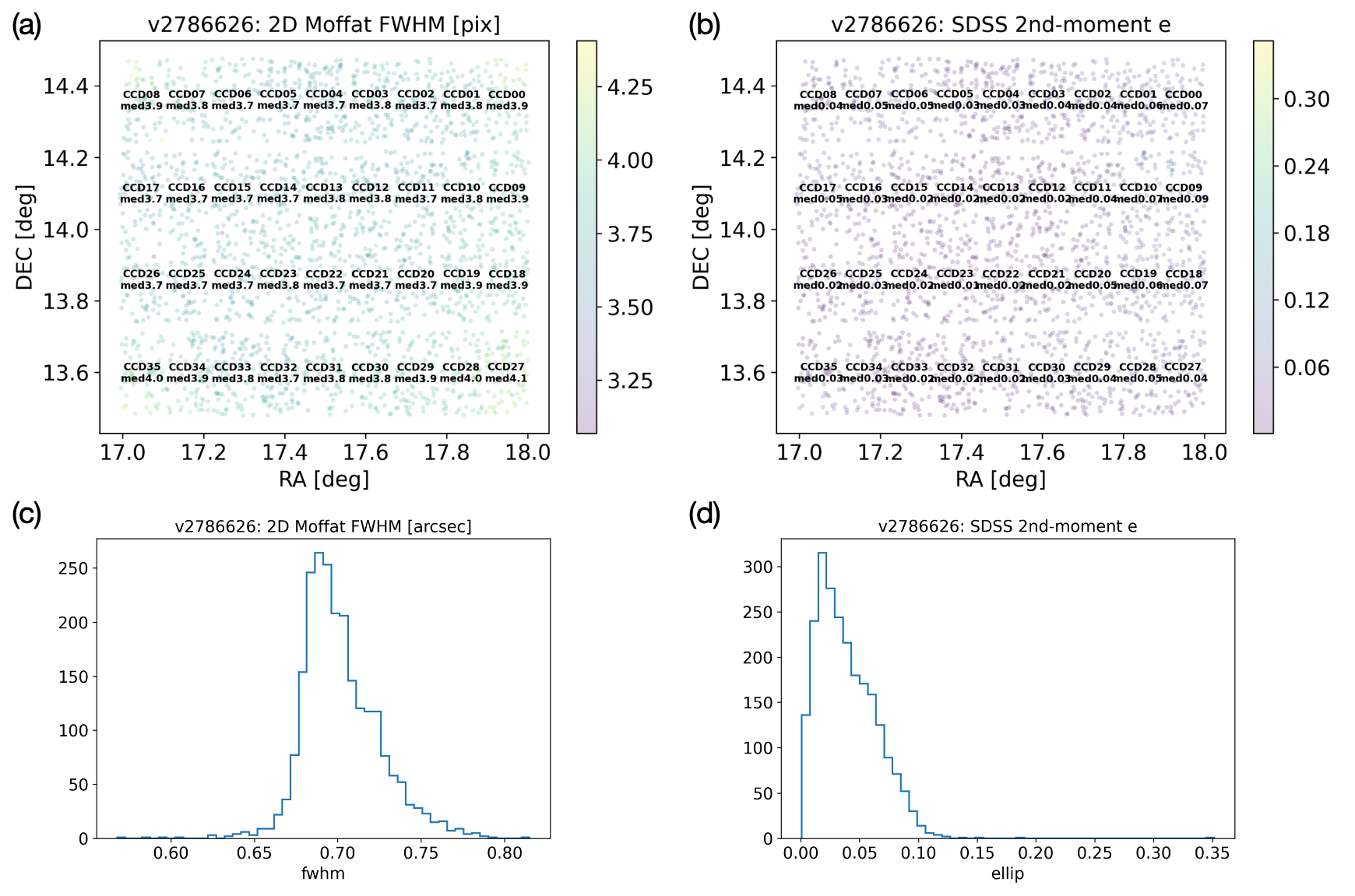}
    \includegraphics[width=0.9\textwidth]{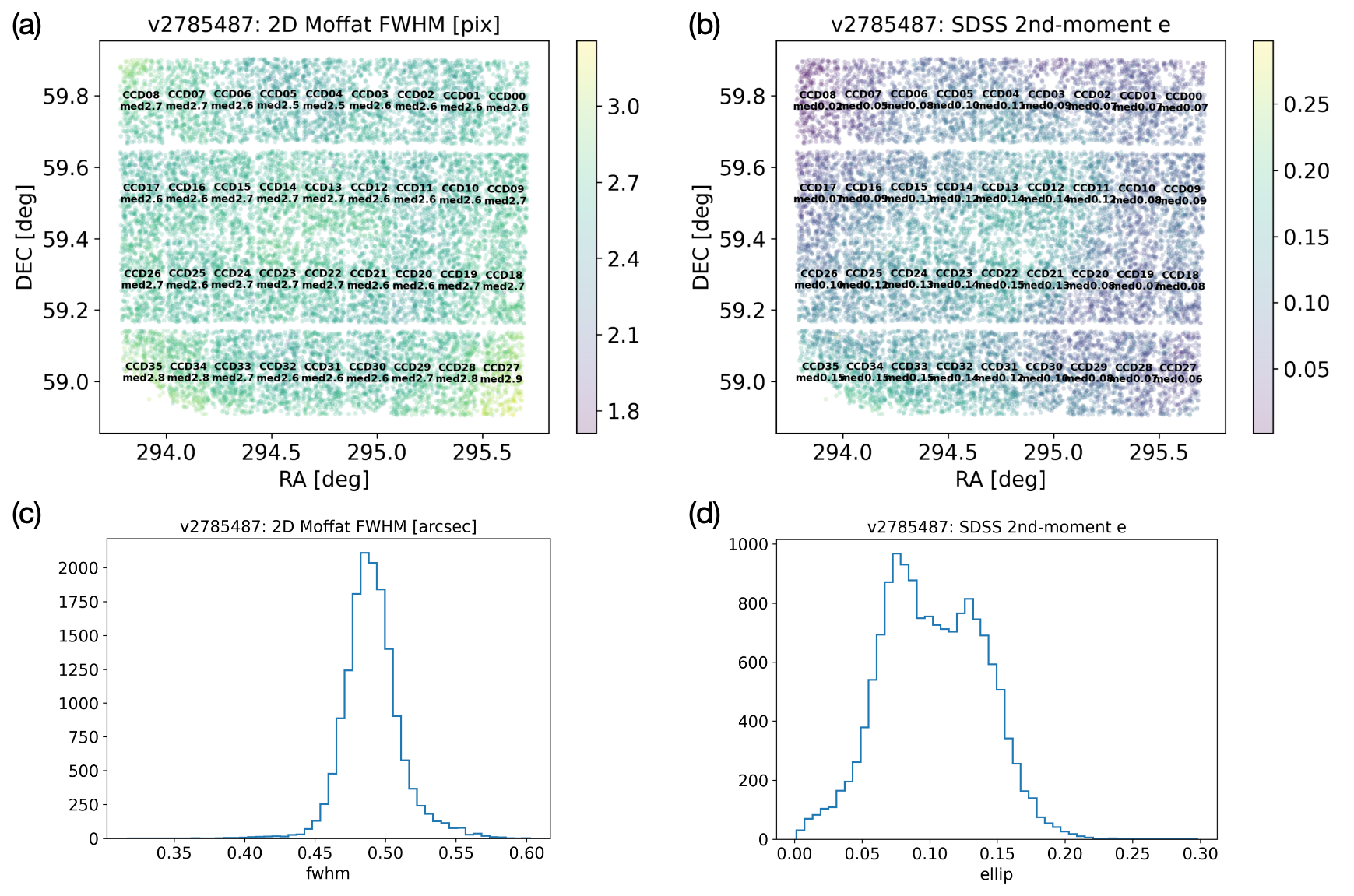}
    \caption{PSF QC for visit 2786626 and 2785487.}
    \label{fig:qc_cluster_2786626}
\end{figure*}

\begin{figure*}
    \centering
    \includegraphics[width=0.9\textwidth]{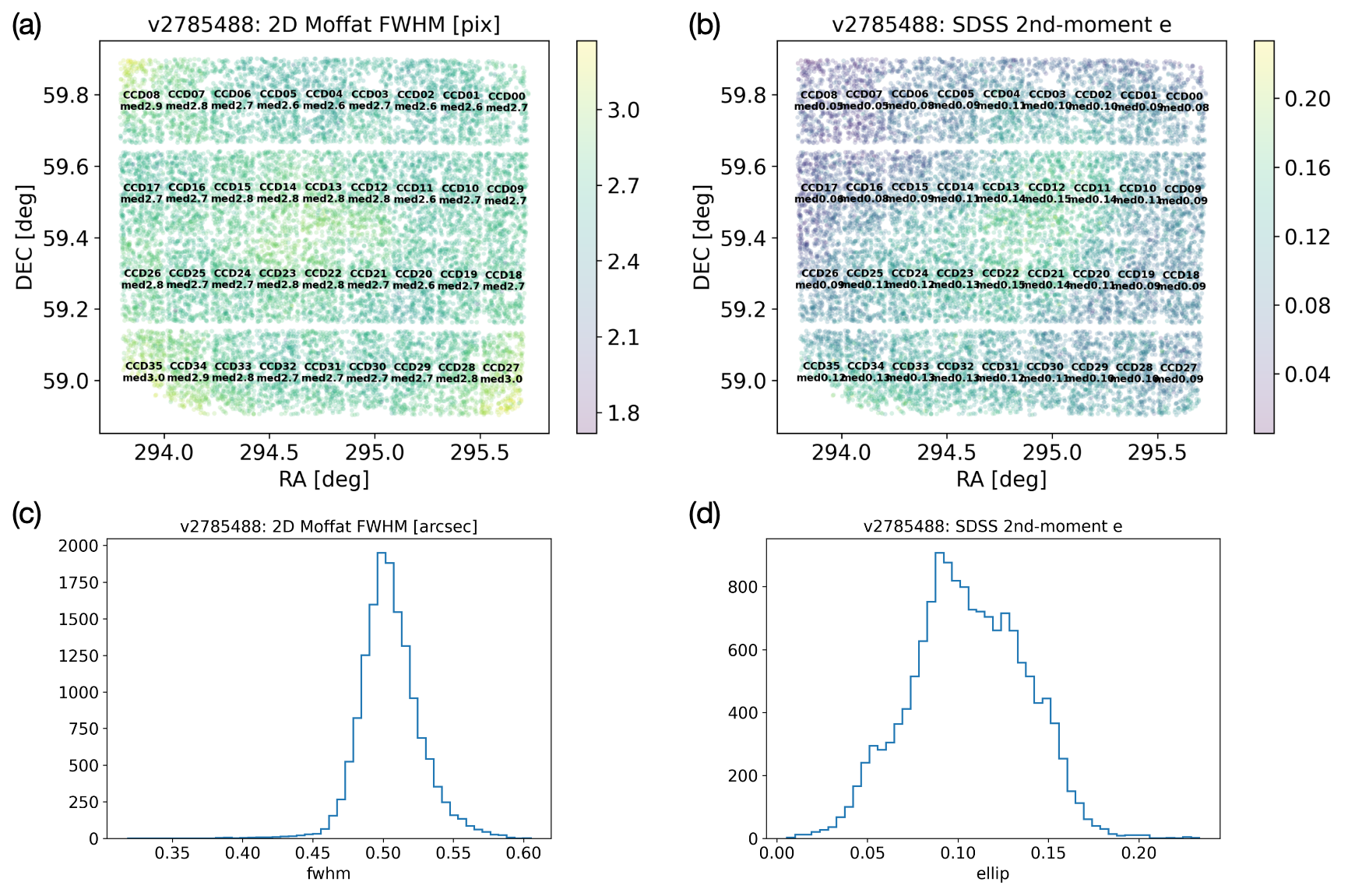}
    \includegraphics[width=0.9\textwidth]{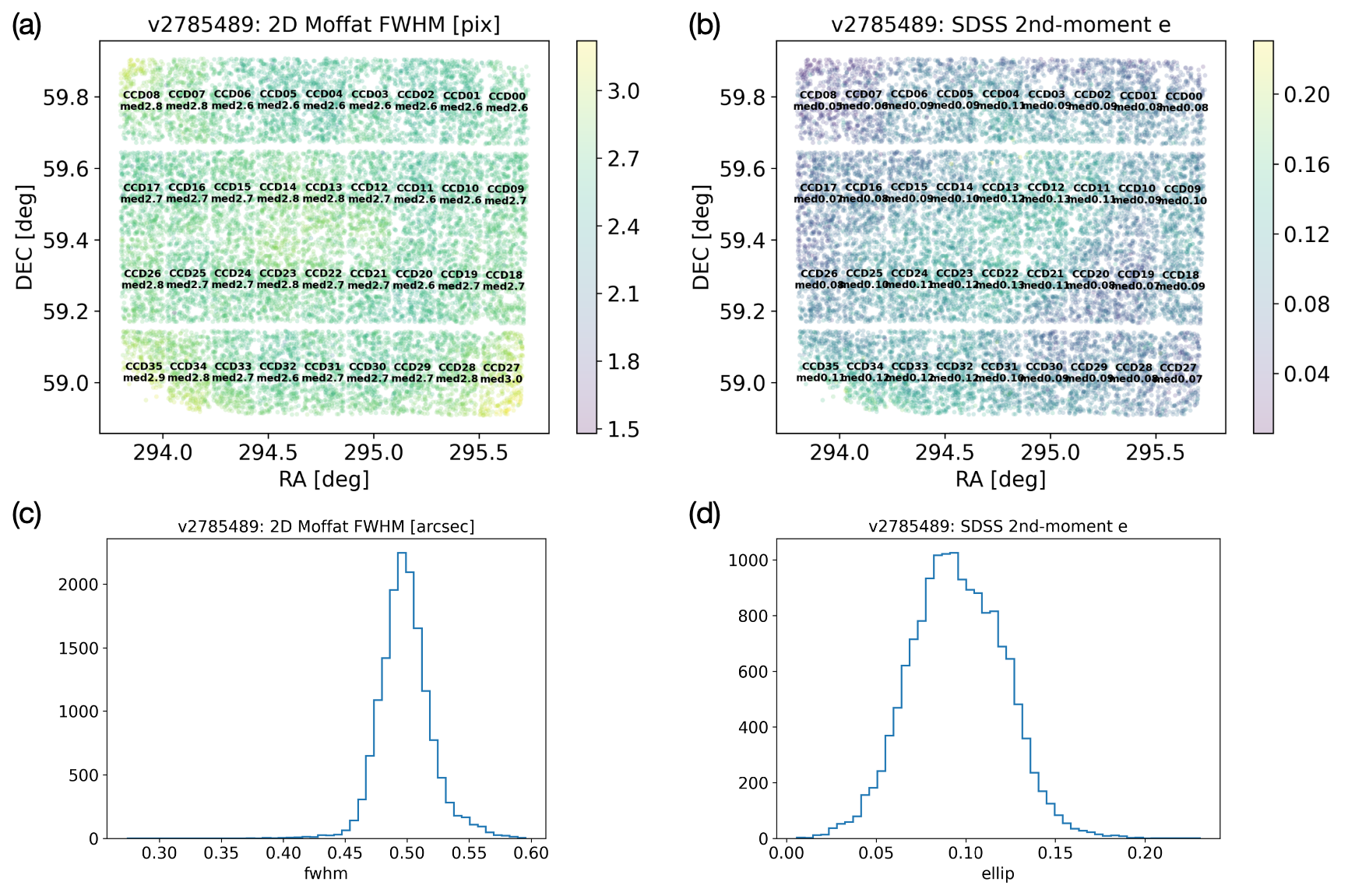}
    \caption{PSF QC for visit 2785488 and 2785489.}
    \label{fig:qc_cluster_2785488}
\end{figure*}

\end{document}